\documentclass[prd,preprint,eqsecnum,nofootinbib,amsmath,amssymb,
               tightenlines,dvips]{revtex4}
 
\usepackage{url}
\usepackage[dvips]{graphicx}

\newcommand{\gton}{\mathrel{\lower.9ex \hbox{$\stackrel{\displaystyle 
>}{\sim}$}}} 
\newcommand{\lton}{\mathrel{\lower.9ex \hbox{$\stackrel{\displaystyle 
<}{\sim}$}}}  

\begin{document}
                                                                              
\vspace*{1cm}
%\preprint{UW/PT 02-21}
                                                                              
\title{QCD Equation of State and Hadron Resonance Gas}
% Or Something Like That
                                                                              
\author{Pasi Huovinen}
\affiliation
    {%
    Institut f\"ur Theoretische Physik,
Johann Wolfgang Goethe-Universit\"at,\\
Max-von-Laue-Stra{\ss}e 1,
60438 Frankfurt am Main, Germany 
    }%
\author{P\'eter Petreczky}
\affiliation
    { Physics Department and RIKEN-BNL Research Center, Brookhaven National Laboratory, Upton, NY 11973, USA
    }%
\date{\today}
                                                                              
\begin{abstract}
We compare the trace anomaly, strangeness and baryon number 
fluctuations  calculated in lattice QCD
with expectations based on hadron resonance gas model.
We find that there is a significant discrepancy between the
hadron resonance gas and the lattice data. This discrepancy is
largely reduced if the hadron spectrum is modified to take
into account the larger values of the quark mass used in lattice
calculations as well as the finite lattice spacing errors.
We also give a simple parametrization of QCD equation of state,
which combines hadron resonance gas at low temperatures with lattice
QCD at high temperatures. We compare this parametrization with
other parametrizations of the equation of state used in hydrodynamical
models and discuss differences in hydrodynamic flow for different equations
of state.

\end{abstract}

\maketitle

\section {Introduction}
Equation of state of hot strongly interacting matter play important
role in cosmology \cite{hindmarsh,laine} and hydrodynamic description
of heavy ion collisions \cite{kolb}. In many cases hydrodynamical
models which try to describe the collective flow in heavy ion
collisions used equation of state (EoS) with first order phase
transition, although lattice QCD shows that the transition to the
deconfined phase is only a crossover \cite{nature}.  It is not obvious
to what extent the collective flow is sensitive to details of the
equation of state (EoS). It turns out, however, that in ideal
  fluid dynamics the anisotropy of the proton flow is particularly
sensitive to the QCD equation of state \cite{pasi05} and using
lattice inspired EoS with crossover transition overpredicts
proton elliptic flow.

Attempts to calculate EoS on the lattice have been made over the last
20 years (see Refs. \cite{me,carleton} for reviews). One of the difficulties in calculating
EoS on the lattice is its sensitivity to high momentum modes and thus to the effects
of finite lattice spacing. This problem is particularly severe in the high temperature
limit. Therefore in recent years calculations have been done using improved staggered
fermions with higher order discretization of the lattice Dirac operators \cite{asqtad_eos,p4_eos}
which largely reduce the cutoff dependence in the high temperature region.
However, there is another source of discretization effects if staggered quarks are used.
The staggered fermion formulation does not preserve the flavor symmetry of continuum QCD. Because of
this the spectrum of low lying hadronic states is distorted and this may effect thermodynamic
quantities in the low temperature region.

Hadron resonance gas (HRG) turned out to be very successful in
describing particle abundances produced in heavy ion collisions
\cite{PBM}. It was also used to estimate QCD transport coefficients \cite{jaki} 
as well as chemical equilibration rates \cite{johanna} close to the
transition temperature.
Thermodynamic quantities calculated in lattice QCD with
rather large quark mass agree well with the HRG model if the masses of the
hadrons in the model are tuned appropriately to match the large quark
mass used in lattice calculations \cite{redlich1}.  Furthermore, the
ratio of certain charge susceptibilities are not very sensitive to the
details of the hadron spectrum and the lattice calculations of these
ratios show a reasonably good agreement with HRG model predictions at
low temperatures \cite{redlich2,redlich3,fluctuations}.
The purpose of this paper is to confront the results of recent lattice
calculations performed with light quark masses with the prediction of
the HRG model and clarify its range of applicability. As we will see
the HRG model describes thermodynamic quantities quite well up to
unexpectedly high temperatures.  Therefore lattice EoS can be combined
with HRG EoS at low temperatures to get rid of large discretization
effects. Such an EoS is also useful for hydrodynamic modeling, and we
construct a parametrization for an EoS interpolating between HRG at
low temperatures and lattice QCD at high temperatures. The rest of the
paper is organized as follows.  In section II we discuss the hadron
resonance gas model and the effects of finite lattice spacing on
hadron masses. Section III deals with the comparison of fluctuations
of baryon number and strangeness with the prediction of HRG model. In
section IV we compare the HRG with the lattice results on trace
anomaly and construct a parametrization of equation of state which is
easy to use in hydrodynamic simulations. In this section we also
  give a detailed comparison with other parametrizations of EoS 
  in the literature.
In section V we discuss hydrodynamic flow
  for different EoSs discussed in the previous section, and highlight
  the differences. Finally section VI contains our conclusions.  In
  appendices we discuss some technical aspects of the calculations, in
  particular the fits of the lattice data as well as the hydrodynamic
  flow for partial chemical equilibrium.

\section{Hadron resonance gas and lattice QCD}

At sufficiently low temperature thermodynamics of strongly interacting matter
at zero baryon number density is dominated by pions. The interaction between pions
is suppressed and chiral perturbation theory can be used to estimate the pion
contribution to thermodynamic potential~\cite{gerber}. In fact, for temperatures $T \le 150$ MeV,
the energy density of pions calculated in 3-loop chiral perturbation theory differs only by less
than $15\%$ from the ideal gas value \cite{gerber}.
As temperature increases heavier hadrons start to contribute to thermodynamics. For temperatures
$T\ge 150$MeV heavy states dominate the energy density. However, the densities of heavy particles
are still small, $n_i \sim \exp(-M_i/T)$, and their mutual interactions, being proportional
to $n_i n_k \sim \exp(-(M_i+M_k)/T)$, are suppressed. Therefore one can use the virial expansion
to evaluate the effect of interactions \cite{dashen}. 
In the low temperature limit the virial expansion reduces to chiral perturbation theory \cite{gerber}.
The virial expansion together with experimental information
on the scattering phase shifts was used by Prakash and Venugopalan to study thermodynamics of 
low temperature
hadronic matter \cite{prakash}. Their analysis showed that there is an interplay between
attractive interactions 
(characterized by positive phase shifts) and repulsive interactions (characterized by negative phase shifts)
such that their net effect can be well approximated by inclusion of free resonances : $\rho$, $K^{*}$, 
$\Delta(1234)$ etc. Thus the interacting gas of hadrons can be fairly well  approximated by 
a non-interacting gas of
resonances corroborating earlier ideas of the statistical bootstrap model \cite{hagedorn}.
To summarize, the partition function of strongly interacting matter at low temperature
can be well approximated by the partition function of non-interacting hadrons and resonances
\begin{eqnarray}
p^{HRG}/T^4 \hspace{-2mm}
&=&\frac{1}{VT^3}\sum_{i\in\;mesons}\hspace{-3mm} 
\ln{\cal Z}^{M}_{m_i}(T,V,\mu_{X^a})
\nonumber \\
&&\hspace{-3mm} +\frac{1}{VT^3}
\sum_{i\in\;baryons}\hspace{-3mm} \ln{\cal Z}^{B}_{m_i}(T,V,\mu_{X^a})\; ,
\label{eq:ZHRG}
\end{eqnarray}
where
\begin{equation}
\ln{\cal Z}^{M/B}_{m_i}
=\mp \frac{V d_i}{{2\pi^2}} \int_0^\infty dk k^2
\ln(1\mp z_ie^{-\varepsilon_i/T}) \quad ,
\label{eq:ZMB}
\end{equation}
with energies $\varepsilon_i=\sqrt{k^2+m_i^2}$, degeneracy 
factors $d_i$ and fugacities
\begin{equation}
z_i=\exp\left((\sum_a X_i^a\mu_{X^a})/T\right) \; .
\label{eq:fuga}
\end{equation}
Here we consider all possible conserved charges $X^a$, including the
baryon number $B$, electric charge $Q$, strangeness $S$ etc.  We do
not include any repulsive interactions in the form of excluded volume
corrections or repulsive mean field~\cite{Satarov}.

The assumption that thermodynamics in the low temperature region is
well described by a gas of non-interacting hadrons and resonances is
important for practical applications of hydrodynamic
models. At the end of the hydrodynamical evolution, the fluid is usually
  converted into particles using the Cooper-Frye
  procedure~\cite{Cooper-Frye}. This procedure conserves energy,
  momentum and charge without any specific considerations if, and only
  if, the equation of state of the fluid is the same than the equation
  of state of free particles~\cite{Laszlo}.
Therefore it is important to confront the predictions of the hadron
resonance gas for different thermodynamical quantities with the
available lattice data.

The most extensive lattice calculations of the equation of state have
been performed using two versions of improved staggered fermions: the
so-called asqtad and p4 formulations \cite{asqtad_eos,p4_eos,hot_eos}.
Calculations have been performed using lattices with temporal extent
$N_{\tau}=4$ and $6$ \cite{asqtad_eos,p4_eos} and more recently with
temporal extent $N_{\tau}=8$ \cite{hot_eos}.  These correspond to
lattice spacings $a=1/(4T)$, $1/(6T)$ and $1/(8T)$ respectively. The
strange quark mass $m_s$ was close to its physical value, while the
light ($u$ and $d$) quark masses were one tenth of the strange quark
mass, $m_q=0.1m_s$, corresponding to the pion mass in the range
$220-260$ MeV.  The lattice calculations of the equation of state have
been compared to the prediction of the hadron resonance gas
\cite{p4_eos,hot_eos}. It turned out that the lattice results fall
considerably below the HRG prediction.  One obvious reason for this
discrepancy is the fact that the quark mass used in lattice
calculations is about a factor two larger then the physical
one. However, this fact alone is unlikely to explain the whole
discrepancy as the contribution of the pseudo-scalar mesons to the
energy density is small and quark mass dependence of other hadron
masses in this small quark mass region is relatively weak.  On the
other hand the lattice spacing dependence of the hadron masses may
play an important role. Since the lattice calculations of the EoS are
performed at fixed temporal extent $N_{\tau}$, the temperature is
varied by changing the lattice spacing $T=1/(N_{\tau}a)$. As the
temperature is decreased the lattice spacing gets larger and the
cutoff effects on the hadron masses increase, i.e. the size of cutoff
effects on the hadron masses is a function of the temperature.

The hadron masses for asqtad improved
staggered fermions have been studied in
Refs. \cite{bernard01,bernard04,bernard07,bazavov09} for several
lattice spacings $a \simeq 0.06$, $0.09$, $0.125$, $0.18$ and $0.25$
fm.  For all of these lattice spacings there are significant deviations
in the hadron masses from the experimental values. Of course, after the
proper continuum extrapolations all the hadron masses agree with the
experiment \cite{bazavov09}.  The large cutoff dependence of the
hadron masses in the staggered formulation is due to large ${\cal
  O}(\alpha_s a^2)$ discretization errors which also break the flavor
symmetry. This is not the case for the improved Wilson fermion formulation
\cite{Durr:2008a,Durr:2008b}, where cutoff dependence of the hadron
masses is below $5\%$ already at lattice spacing $<0.2$fm.  In the
following subsections we discuss the cutoff dependence of the
pseudo-scalar meson, vector meson and baryon masses separately.

\subsection{Pseudo-scalar mesons in staggered formulation}
\label{sec:ps}
The staggered formulation of lattice QCD describes four degenerate quark flavors
in the continuum limit. To obtain the physical number of flavors, i.e.~one relatively
heavy strange quark and two light quarks, the so-called rooting procedure is used.
The rooting procedure amounts to replacing the fermion determinant in the path integral
expression of the partition function with its fourth root. 
In the continuum limit this procedure is justified
\cite{continuum_root}\footnote{The justification of the rooting
  procedure at finite lattice spacing is still subject of debate see
  e.g. Ref. \cite{creutz07,kronfeld07}}. At finite lattice spacing,
however, the four flavors are not degenerate, but there are flavor
changing discretization effects of order ${\cal O}(\alpha_s a^2)$.
As the result of this the sixteen pseudo-scalar mesons of the 4 flavor theory have unequal masses,
and only one of them has a vanishing mass in the limit of the zero quark mass, $m_q \rightarrow 0$. 
The sixteen pseudo-scalar mesons can be grouped into eight multiplets \cite{golterman86}.
The multiplets are characterized
by the different masses $m_{{\rm ps}_i}$ and degeneracies $d_{\rm ps}^i$, 
$i=0,1,..7$. 
The first multiplet contains one
Goldstone pseudo-scalar meson, i.e.  $m_{{\rm ps}_0}^2 \sim m_q$ and $d_{\rm ps}^0=1$.
The flavor generators $\Gamma^F$ of the 4 flavor theory can be chosen to be the product of the Dirac
matrices \cite{golterman86,kluberg}.
The masses of other pseudo-scalar mesons are given by 
\begin{equation}
m_{{\rm ps}_i}^2=m_{{\rm ps}_0}^2 + \delta m_{{\rm ps}_i}^2.
\label{m_ps}
\end{equation}
The quadratic splittings $\delta m_{ps_i}^2$ are independent of the input quark mass 
$m_q$ to a very good approximation
and are proportional to $(\alpha_s a^2)$ for small lattice spacings, and in general, increase with increasing
index $i$. The correspondence between the index $i$ and the flavor matrix as well as 
the degeneracies $d_{\rm ps}^i$ for non-Goldstone pseudo-scalar mesons are given
in Table \ref{tab:ps_par}.

\begin{figure}
\includegraphics[width=10cm]{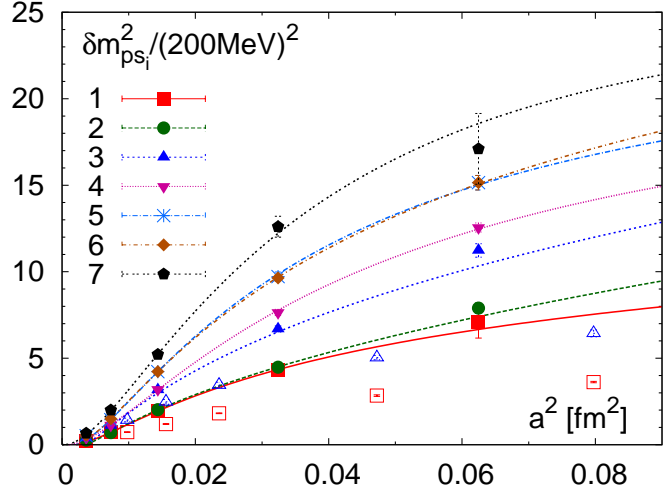}
\caption{The quadratic  splittings of non-Goldstone pseudo-scalar mesons in the seven different
multiplets calculated with asqtad action \cite{bazavov09}
at different lattice spacings. The lines show the parametrization given by Eq. (\ref{delta_ps}).
The open symbols refer to the lattice data obtained with the stout action \cite{BW_Tc_09}.}
\label{fig:pion}
\end{figure}

The quadratic splittings have been calculated in numerical simulations with asqtad action 
in Refs.~\cite{bernard04,bazavov09}.
For lattice spacings $0.09 {\rm fm} <a< 0.25 {\rm fm}$  the data can be  well parametrized 
by the form 
\begin{equation}
r_1^2 \cdot \delta m_{ps_i}^2=\frac{a_{\rm ps}^i x+b_{\rm ps}^i x^2}{{(1+c_{\rm ps}^i x)}^{\beta_i}},
~x=(a/r_1)^2.
\label{delta_ps}
\end{equation} 
Here and in what follows we use the scale parameter
$r_1$ extracted from the static quark potential $V(r)$  to convert from lattice units to physical units.
The scale parameter $r_1$ is defined as 
\begin{equation}
r^2 \frac{d V(r)}{d r}|_{r=r_1}=1.0.
\end{equation}
We use the value $r_1=0.318$ fm determined from bottomonium splitting \cite{bernard04}.
The values of the parameters 
$a_{\rm ps}^i$,  $b_{\rm ps}^i$, $c_{\rm ps}^i$, $\beta_{\rm ps}^i$ are given in Table \ref{tab:ps_par}.
\begin{table}
\begin{tabular}{|c|c|c|c|c|c|c|}
\hline
i    & $\Gamma^F$ &
$d_{\rm ps}^i$ & $a_{\rm ps}^i$ &  $b_{\rm ps}^i$ & $c_{\rm ps}^i$ & $\beta_{\rm ps}^i$ \\
\hline
1  &  $\gamma_0 \gamma_5 $ &  1   & 7.96583        &   45.6265       &  -0.983624     & 1.80  \\
2  &  $\gamma_i \gamma_5 $ &  3   & 25.9514        &   129.049       &  -4.673780     & 1.45  \\
3  &  $\gamma_i \gamma_j $ &  3   & 19.3047        &   163.787       &  -5.675470     & 1.55  \\
4  &  $\gamma_i \gamma_0 $ &  3   & 4.26042        &   45.0193       &  -0.489748     & 2.00  \\
5  &  $\gamma_i $          &  3   & 5.43308        &   79.455        &  -1.71669      & 2.00  \\
6  &  $\gamma_0 $          &  1   & 7.52963        &   95.2536       &  -2.24599      & 1.80  \\
7  &  $1$                  &  1   & 3.76433        &   70.8311       &  -0.373003     & 2.20  \\
\hline
\end{tabular}
\caption{The parameters of Eq. (\ref{delta_ps}) describing the quadratic pseudo-scalar 
meson splittings. Also shown are the flavor matrices $\Gamma^F$ and the degeneracy factors $d_{\rm ps}^i$.}
\label{tab:ps_par}
\end{table}
In Figure \ref{fig:pion} we show the quadratic pseudo-scalar meson 
splittings calculated by the MILC collaboration~\cite{bernard04,bazavov09} as function of the lattice spacing
and compared with the parametrization given by Eq.~(\ref{delta_ps}). As one can see from the Figure~\ref{fig:pion},
this parametrization gives good description of the data. 
In Figure \ref{fig:pion} we also show the pseudo-scalar
meson splitting for stout action used by the Budapest-Wuppertal group \cite{BW_Tc_09}. The splittings
are significantly reduced compared to the calculations with asqtad action.
To take into account the flavor symmetry breaking in the pseudo-scalar meson sector, i.e.~the fact
that pion masses are non-degenerate the contributions of pions and kaons to the pressure is
calculated as 
\begin{equation}
p^{\pi,K}/T^4=\frac{1}{16} \frac{1}{VT^3} \sum_{i=0}^7 d_{\rm ps}^i \ln{\cal Z}^{M}(m_{{\rm ps}_i}),
\label{ps_contr}
\end{equation} 
where $m_{{\rm ps}_i},~i=1-7$ is calculated according to Eq. (\ref{m_ps}) and $m_{{\rm ps}_0}$ is
equal to the pion or kaon mass used in the actual lattice calculations. Figure \ref{fig:pion} shows
that the splitting between different pseudo-scalar mesons is quite large for lattice spacings used in
calculations of the equations of state ($a=0.12-0.18$)fm. Even in the calculations with the stout action
the mass of the heaviest pion that enters Eq. (\ref{ps_contr}) is about $500$MeV if $N_{\tau}=8$ lattices are used.
As the result the contribution of the pseudo-scalar mesons to thermodynamic quantities at finite lattice
spacings is smaller than in the continuum.

\subsection{Vector meson masses}
In lattice calculations hadron masses are functions of the input quark mass and the lattice
spacing. The quark mass dependence of the hadron masses is usually studied in terms of the 
lightest (Goldstone) pion mass. Since in lattice calculations quark masses are larger than the
physical ones, 
and there is no full flavored chiral symmetry 
at finite lattice spacing 
a combined extrapolation in the pion mass and lattice spacing is needed for all hadron masses.
We fitted the lattice spacing and pion mass dependence of the vector meson masses with
the following simple formula
\begin{equation}
r_1 \cdot m(a,m_{\pi})=r_1 m^{phys}+\frac{a_1 (r_1 m_{\pi})^2-a_1 (r_1 m_{\pi}^{phys})^2}{1+a_2 x}+
\frac{b_1 x}{1+b_2 x},~x=(a/r_1)^2,~
\label{mV}
\end{equation}
where $m^{phys}$ is the physical value of the meson mass from the particle data book.
It turns out that this formula describes the lattice spacing dependence of the vector meson
masses in the interval $0.06 {\rm fm} <a <0.18 {\rm fm}$ for $(r_1 m_{\pi})^2<0.8$. 
For strange vector mesons we set $a_2=0$ as there is no apparent lattice spacing
dependence of the slope in their quark mass dependence.
The numerical values of $a_1,~a_2,~b_1$ and $b_2$ are given in the Appendix A, where also
the lattice data on the vector meson masses are shown.

\subsection{Baryon masses}
The nucleon and the $\Omega$ baryon masses have been calculated 
by the MILC collaboration with asqtad action at five different
lattice spacings, $a=0.06, 0.09, 0.12$ and $0.18$fm 
\cite{bernard01,bernard04,bazavov09,milc_unpub} and several quark masses.
We performed a simultaneous fit of their quark mass (pion mass) and
the lattice spacing dependence  using the  Ansatz given by Eq. (\ref{mV}),
which works well for $(r_1 m_{\pi})^2<0.8$. The parameters of this fit together with the lattice
data on the nucleon and $\Omega$ baryon masses are presented in the Appendix A. 
In calculations with asqtad action the value of the
strange quark mass was slightly larger than its physical value for each lattice spacing
considered. This is due to the fact that the strange quark mass was fixed considering
the ratio of the $\phi$ meson mass to the mass of the unmixed $\eta_{ss}$ pseudo-scalar meson 
instead fixing the kaon mass. This gives difference ${\cal O}(a^2)$ in the value of the strange
quark mass. To take this into account the lattice spacing dependence of the strange quark mass
was parametrized as $m_s/m_s^{phys}(a)=1+1.02(a/r_1)^2$. Then to estimate the $\Omega$ baryon
mass we used the following formula
\begin{equation}
r_1 m_{\Omega}(a,m_{\pi})=r_1 m_{\Omega}^{phys}+a_1 (r_1 m_{\pi})^2-a_1 (r_1 m_{\pi}^{phys})^2 +
b_1 x+(m_{\Omega}^{phys}-m_{\Delta}^{phys})\cdot 1.02 x,~x=(a/r_1)^2.
\label{mOmega}
\end{equation}
Here the last term accounts for  the small
deviation of the strange quark mass from its physical value.
For other baryons ($\Delta$, $\Lambda$, $\Sigma$, 
and $\Xi$) no such detailed lattice calculations are available. 
Therefore to estimate the cutoff dependence of the $\Delta$ mass
we use the same formula as for nucleon. While to estimate the cutoff
dependence of the masses of the strange baryons
we used the following formulas
\begin{eqnarray}
&\displaystyle
r_1 \cdot m_{\Lambda}(a,m_{\pi})=m_{\Lambda}^{phys}+\frac{2}{3}\frac{a_1 (r_1 m_{\pi})^2}{1+a_2 x}+
\frac{b_1 x}{1+b_2 x}+\frac{r_1 \cdot ( m_{\Lambda}^{phys}-m_{p}^{phys})}{1+a_2 x} 
\left( \frac{m_s}{m_s^{phys}} \right),\label{mLambda} \\
&\displaystyle
r_1 \cdot m_{\Sigma}(a,m_{\pi})=m_{\Sigma}^{phys}+\frac{1}{3}\frac{a_1 (r_1 m_{\pi})^2}{1+a_2 x}+
\frac{b_1 x}{1+b_2 x}+\frac{r_1 \cdot ( m_{\Sigma}^{phys}-m_{p}^{phys})}{1+a_2 x} 
\left( \frac{m_s}{m_s^{phys}} \right), \label{mSigma}\\
&\displaystyle
r_1 \cdot m_{\Xi}(a,m_{\pi})=m_{\Xi}^{phys}+\frac{1}{3}\frac{a_1 (r_1 m_{\pi})^2}{1+a_2 x}+
\frac{b_1 x}{1+b_2 x}+\frac{r_1 \cdot (m_{\Xi}^{phys}-m_{p}^{phys})}{1+a_2 x}
\left( \frac{m_s}{m_s^{phys}} \right)\label{mXi} \\
&\displaystyle
x=(a/r_1)^2. \nonumber
\end{eqnarray}
Here again we have taken into account that the strange quark mass in simulations
with asqtad was slightly larger than the physical value. 
In the Appendix we give the comparison of the baryon masses calculated using the above
formulas with available lattice results. It turns out that our parametrization of the baryon
masses works reasonably well. We will use these parametrizations when calculating different
quantities in the HRG model in the following sections.

\section{Fluctuations of conserved charges}

Derivatives of the pressure with respect to chemical potential of conserved charges, 
e.g. baryon number ($B$), electric charge $Q$ and strangeness $S$ can be easily calculated in lattice
QCD
\begin{equation}
\chi_n^X=T^n \frac{\partial^n p(T,\mu_B,\mu_Q, \mu_S)}{\partial \mu_X^n}|_{\mu_X=0},~X=B,Q,S.
\end{equation}
These are related to quadratic and higher order fluctuations 
of conserved charges $\chi_2^X=\langle X^2 \rangle/(V T^3)$, 
$\chi_4^X=( \langle N_X^4 \rangle - 3 \langle N_X^2 \rangle^2)/(VT^3)$
etc.\footnote{Here we consider the case of zero chemical potential, so $\langle N_X \rangle=0$.} 
Contrary to the pressure itself the evaluation of these derivatives does not involve zero
temperature subtraction and integration in the temperature variable starting from some low
temperature value. Therefore it is easy to compare them to the prediction of the HRG model.
Different fluctuations up to the sixth order have been calculated for p4 and asqtad action
\cite{fluctuations,milcTc,milc_lat09} on $N_{\tau}=4$ and $N_{\tau}=6$ lattices.
Quadratic strangeness fluctuations have been also calculated on $N_{\tau}=8$ 
lattices for the p4 and asqtad actions by the HotQCD collaboration \cite{hot_eos}.
While there are extensive calculations with p4 action at finite temperature, the zero temperature
hadron spectrum was not studied in detail for p4 action. Therefore our analysis of fluctuations
and comparison with HRG model will mostly rely on results obtained with asqtad action.

\begin{figure}
\includegraphics[width=9cm]{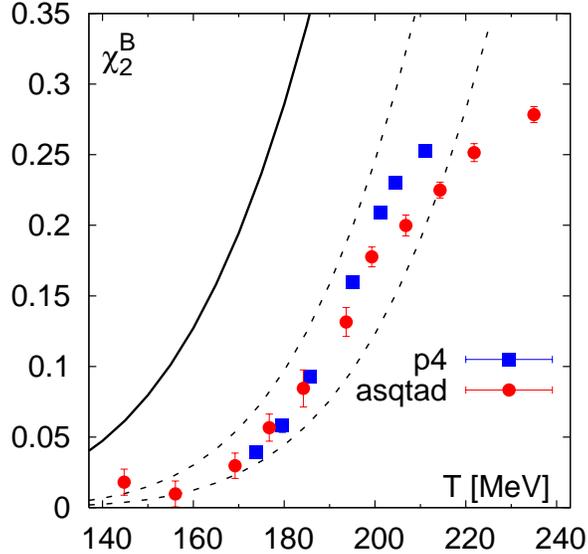}
\caption{Baryon number fluctuations calculated with asqtad action on the $N_{\tau}=6$ lattices
compared with HRG model with physical value of the baryon masses (solid line) and with HRG model with 
baryon masses calculated according Eqs. (\ref{mV})-(\ref{mXi}), $m_{cut}^B=1.8$GeV and $m_{cut}^B=2.5$GeV 
(dashed lines). Also shown are the lattice results for the p4 action.}
\label{fig:chiB}
\end{figure}

Let us start our discussion with baryon number fluctuations. Baryon number fluctuations
for asqtad action have been calculated in Ref.~\cite{milcTc} for $m_q=0.2m_s$ on $N_{\tau}=6$
lattices. Equations (\ref{mV})-(\ref{mXi}) describe the quark mass and lattice spacing dependence
of ground state baryon masses calculated on lattice with asqtad action. Nothing is known about
the lattice spacing dependence of the excited baryon masses which play very important role in baryon
number fluctuations in the temperature range of interest. We assume that all the excited baryons 
up to certain mass threshold $m_{cut}^B$ have
the same quark mass and lattice spacing dependence, while the baryon masses above that threshold are not
modified by finite lattice spacing.
The mass threshold $m_{cut}^B$ is an additional
parameter of our model. In Fig. \ref{fig:chiB} we show the lattice data for baryon number fluctuations
for asqtad action compared with hadron resonance gas model with physical value of the baryon masses
including all the states up to $2.5$GeV. The lattice data fall considerably below the HRG prediction.
The baryon number fluctuations have also been calculated in 
a HRG model, where all the baryon masses up to the $m_{cut}^B=1.8$GeV and $m_{cut}^B=2.5$GeV
have been modified according to Eqs.(\ref{mV})-(\ref{mXi}) . The corresponding results are shown
as dashed lines. As one can see,  the HRG overshoots the lattice data with $m_{cut}^B=1.8$GeV, while with
$m_{cut}^B=2.5$GeV it undershoots the lattice data. 
However, the agreement between lattice data and HRG is greatly improved.
For completeness we also show the lattice data for p4
action calculated for $m_q=0.1m_s$ \cite{fluctuations}. 

\begin{figure}
\includegraphics[width=8cm]{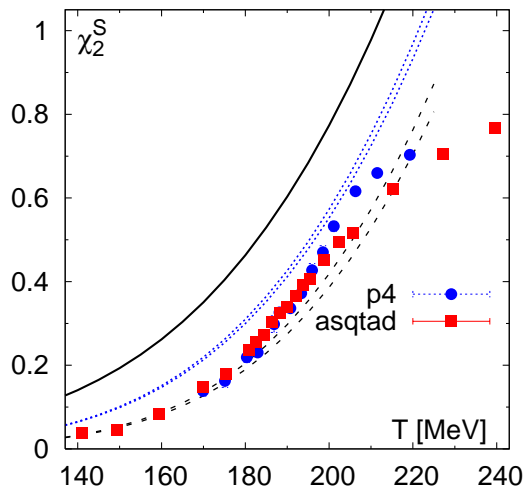}
\caption{The strangeness fluctuations calculated on $N_{\tau}=8$ lattices for asqtad and p4 actions
and compared with the prediction of the HRG model with physical (solid line) and modified (dashed lines)
hadron masses. The upper (lower) dashed line corresponds to $m_{cut}^B=1.8(2.5)$GeV.
The doted lines show the prediction of the HRG with modified hadron masses for $N_{\tau}=12$.
 }
\label{fig:chiS}
\end{figure}

Strangeness fluctuations have also been calculated on the lattice using asqtad and p4 action 
\cite{fluctuations,hot_eos,milcTc,milc_lat09} on $N_{\tau}=6$ and $N_{\tau}=8$ lattices.
We have calculated strangeness fluctuations in HRG model, where ground vector state meson
masses have been calculated using Eq. (\ref{mV}), while baryon masses have been calculated using
Eqs. (\ref{mV})-(\ref{mXi}). The contribution of kaons has been treated in the way discussed in 
section \ref{sec:ps}, i.e.~for each physical kaon averaging over sixteen staggered flavors have been
performed and the mass splitting has been calculated using Eq.~(\ref{delta_ps}). 
This turns out to be 
important for the description of strangeness fluctuation for 
temperatures $T<165$MeV, where the contribution of kaons is quite significant. In Fig. \ref{fig:chiS}
we show the prediction of the HRG model with modified hadron masses compared to the lattice data
for asqtad action for $N_{\tau}=8$. In the figure we show the prediction of the HRG model with
cutoffs $m_{cut}^B=1.8$GeV and $m_{cut}^B=2.5$GeV for the baryon masses. Here the effect of using
different cutoffs for the modification of the baryon masses is significantly smaller as meson
contribution to strangeness fluctuations is large. In the figure we also show the prediction of
the HRG model with physical hadron masses as well as the lattice results for the p4 action.
As one can see the lattice data fall significantly below the predictions of HRG model with physical
quark masses, while the HRG model with modified hadron masses gives quite a good description of the lattice
data. For completeness we also show the result for HRG model with modified hadron masses for $N_{\tau}=12$.
As one can see from Figs.~\ref{fig:chiB} and \ref{fig:chiS}, HRG model can describe baryon number and
strangeness fluctuations reasonably well up to temperatures as high as $T_c$. 

\section{QCD equation of state}

\subsection{The trace anomaly and parametrization of the equation of state}
Available lattice data provide an EoS which is not easy to use
in hydrodynamic models because different
thermodynamic quantities are suppressed in the low temperature region
due to discretization errors.
In the previous section we have seen this for  baryon number and strangeness
fluctuations.  

In lattice QCD the calculation of the pressure, energy density and entropy density
usually proceeds through the calculation of the trace anomaly $\Theta(T)=\epsilon(T) -3 p(T)$. Using
the thermodynamic identity the pressure difference at temperatures $T$ and $T_{\rm low}$ can be expressed
as the integral of the trace anomaly 
\begin{equation}
\frac{p(T)}{T^4}-\frac{p(T_{\rm low})}{T_{\rm low}}=\int_{T_{\rm low}}^{T} \frac{d T'}{{T'}^5} \Theta(T).
 \label{P-integral}
\end{equation}
By choosing the lower integration limit sufficiently small, $p(T_{\rm low})$ can be neglected due
to the exponential suppression. Then the  energy density $\epsilon(T)=\Theta(T)+3 p(T)$ and
the entropy density $s(T)=(\epsilon+p)/T$ can be calculated. This procedure is known as 
the integral method \cite{boyd}.
Since the trace anomaly plays a central role in lattice determination of the equation of state,  we 
will discuss it in the HRG model
and its comparison with the lattice data in the following. As we will see this helps constructing realistic 
equation of state that can be used in hydrodynamic models. 

\begin{figure}
\includegraphics[width=9cm]{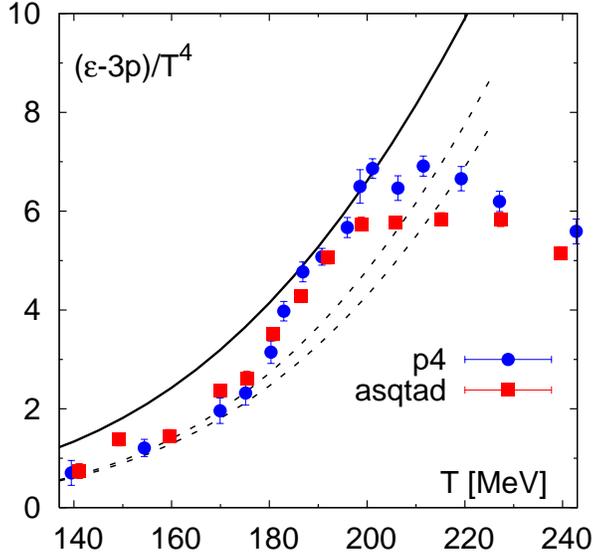}
\caption{The trace anomaly calculated in lattice QCD compared with the HRG model with
physical hadron masses (solid line) and modified hadron masses (dashed lines).
The upper (lower) dashed line corresponds to $m_{cut}^B=1.8(2.5)$GeV. }
\label{fig:e-3p}
\end{figure}

As mentioned before, finite temperature lattice calculations are usually performed at fixed
temporal extent $N_{\tau}$ and the temperature is varied by varying
the lattice spacing $a$, $T=1/(N_{\tau} a)$. Thus, calculations at low
temperatures are performed on coarse lattices, while the lattice
spacing gets smaller as the temperature is increased. Consequently 
the trace anomaly can be accurately
calculated in the high
temperature region, while in the low temperature region it is affected
by possibly large discretization effects. Therefore to construct
realistic equation of state we could use the lattice data for the
trace anomaly in the high temperature region, $T>250$MeV, 
and use
HRG model in the low temperature region, $T\lton 180$ MeV. In
Fig.~\ref{fig:e-3p} we compare the lattice results on trace anomaly
obtained on $N_{\tau}=8$ lattices with asqtad and p4 action with the
HRG model.
The HRG model with modified masses appears to describe the lattice
data quite well up to temperatures of about $180$MeV. In the
intermediate temperature region $180{\rm MeV} \lton T < 250 {\rm MeV}$
the HRG model is no longer reliable, whereas 
discretization effects in lattice calculations
could be large. The later can be seen by comparing the lattice data
obtained on $N_{\tau}=6$ and $N_{\tau}=8$ lattices with p4 and asqtad
action. Therefore we constrain the trace anomaly in the intermediate
region only by the value of the entropy density at high temperatures.

In pure gauge theory, where continuum extrapolation has been performed,
 the entropy density falls below the ideal gas limit only by $15\%$ at
temperatures of about $4T_c$~\cite{boyd} ($T_c$ is the transition temperature). In QCD the entropy density
calculated on $N_{\tau}=6$ and $8$ lattices is $(5-10)\%$ below the ideal gas limit \cite{petr_qm09}
at the highest available temperature. Furthermore, resummed perturbative calculations 
describe quite well the entropy density in the high temperature region both in pure gauge theory
\cite{blaizot} and in QCD \cite{petr_qm09}. These calculations  indicate
that deviation from the ideal gas limit is about $(5-10)\%$. The fluctuations of quark number 
are also close to ideal gas limit and well described by resummed perturbative 
calculations \cite{petr_qm09}. Here the deviation from the ideal gas limit is less than $5\%$.
Therefore we will use the guidance from existing lattice QCD calculations and 
require that the entropy density is below the ideal gas limit by either $5\%$ or $10\%$, 
when parametrizing the trace anomaly.

At high temperature the trace anomaly can be well parametrized by the inverse polynomial form
(see e.g.~Ref.~\cite{hot_eos}). Therefore we will use the following Ansatz for the high temperature
region
\begin{equation}
(e-3P)/T^4 = d_2/T^2 + d_4/T^4 + c_1/T^{n_1} + c_2/T^{n_2}.
\label{e-3p_high}
\end{equation}
This form does not have the right asymptotic behavior in the high
temperature region, where we expect $(e-3P)/T^4 \sim g^4(T)
~1/\ln^2(T/\Lambda_{QCD})$, but works well in the temperature range of
interest. Furthermore, it is flexible enough to do the matching to the
HRG result in low temperature region. We match to the HRG model at
temperature $T_0$ by requiring that the trace anomaly as well as its
first and second derivatives are continuous. The parametrization of
the trace anomaly and thus QCD equation of state obtained using these
requirements are labeled by $s95p\rm{-v1}$, $s95n\rm{-v1}$ and
$s90f\rm{-v1}$. The labels ``$s95$'' and ``$s90$'' refer to the
fraction of the ideal entropy density reached at $T=800$MeV ( 95\% and
90\% respectively), whereas the labels $p$, $n$ and $f$ refer to a
specific treatment of the peak of the trace anomaly or its matching to
the HRG.  The detailed procedure of performing the fit to the lattice
data and matching to the HRG model is described in Appendix B, where also
the labeling scheme is explained in more detail. The values of the
parameters $T_0,~d_2,~d_4,~c_1,~c_2,~n_1$ and $n_2$ in each
parametrization are given in Table~\ref{tab:par}.
\begin{table}
\begin{tabular}{cccccccc}
\hline
         & $d_2$ (GeV$^2$) & $d_4$ (GeV$^4$)       & $c_1$ (GeV$^{n_1}$)    & $c_2$ (GeV$^{n_2}$)     & $n_1$ &  $n_2$  & $T_0$ (MeV) \\   
\hline      
$s95p$   & 0.2660        & 2.403$\cdot 10^{-3}$& -2.809$\cdot 10^{-7}$& 6.073$\cdot 10^{-23}$&  10   &  30     &   183.8   \\
$s95n$   & 0.2654        & 6.563$\cdot 10^{-3}$& -4.370$\cdot 10^{-5}$& 5.774$\cdot 10^{-6}$ &  8    &  9      &   171.8   \\
$s90f$   & 0.2495        & 1.355$\cdot 10^{-2}$& -3.237$\cdot 10^{-3}$& 1.439$\cdot 10^{-14}$&  5    &  18     &   170.0   \\
\hline
\end{tabular}
\caption{The values of the parameters for different fits of the trace anomaly.}
\label{tab:par} 
\end{table} 
The HRG result
for the trace anomaly can also be parametrized by the simple form
\begin{equation}
\frac{\epsilon -3P}{T^4} = a_1 T +  a_2 T^3
                         + a_3 T^4 + a_4 T^{10},
  \label{e-3p_low}
\end{equation}
with $a_1 = 4.654$ GeV$^{-1}$, $a_2 = -879$ GeV$^{-3}$, $a_3 = 8081$
GeV$^{-4}$, $a_4 = -7039000$ GeV$^{-10}$ (see Appendix B for details).

\begin{figure}
\includegraphics[width=10cm]{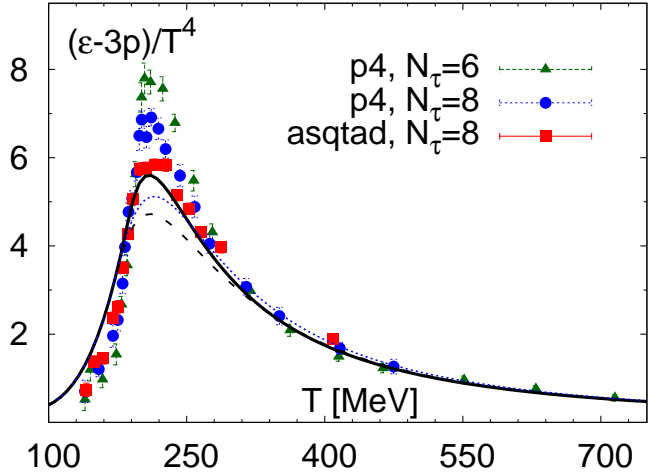}
\caption{The trace anomaly calculated in lattice QCD with p4 and
  asqtad actions on $N_{\tau}=6$ and $8$ lattices compared with the
  parametrization given by Eqs. (\ref{e-3p_high}) and
  (\ref{e-3p_low}).  The solid, dotted and dashed lines correspond to
  parametrizations $s95p{\rm -v1}$, $s95n{\rm -v1}$ and $s90f\rm{-v1}$
  respectively, as discussed in the text.}
\label{fig:e-3p_param}
\end{figure}

The lattice data for trace anomaly compared to the parametrization
given by Eqs. (\ref{e-3p_high}) and (\ref{e-3p_low}) is shown in
Fig.~\ref{fig:e-3p_param}.  We show three parametrizations in the
figure corresponding to a entropy density at $T=800$MeV which is
below the ideal gas limit by $5\%$ ($s95p\rm{-v1}$ and $s95n{\rm-v1}$,
the solid and dotted lines, respectively) and $10\%$ ($s90f\rm{-v1}$,
dashed line). All the parametrizations describe the lattice data for
$T>250$MeV, while in the low temperature region, $T<170$MeV, they are
significantly above the lattice results. On the other hand, our
parametrizations of the trace anomaly are below the lattice data in
the peak region.  This comes from the imposed constraint on the
entropy density at $T=800$MeV. If we would use a parametrization which
goes through the $N_\tau=8$, p4 lattice data and matches the resonance
gas model at some temperature near $190$MeV, the entropy density would
overshoot the ideal gas limit already at temperatures of about
$600-700$MeV.  Such a behavior would contradict the available lattice
and weak coupling results.

The difference in the $s95n\rm{-v1}$ and $s95p{\rm -v1}$ parametrizations is in the
  treatment of the peak region. When we do the fit only on the lattice
  data above $T=250$MeV ($s95n{\rm -v1}$, dotted line), the peak value is
  clearly below the present data. To explore the sensitivity of the
  EoS to the height of the peak, we also did the fit using one
  additional point at $T=205$MeV close to the present data, and the
  same entropy constrain ($s95p{\rm -v1}$, solid line). This forces the trace
  anomaly almost to reach the data at the peak maintaining a
  reasonable fit to the data at high temperatures.
\begin{figure}
\includegraphics[width=8cm]{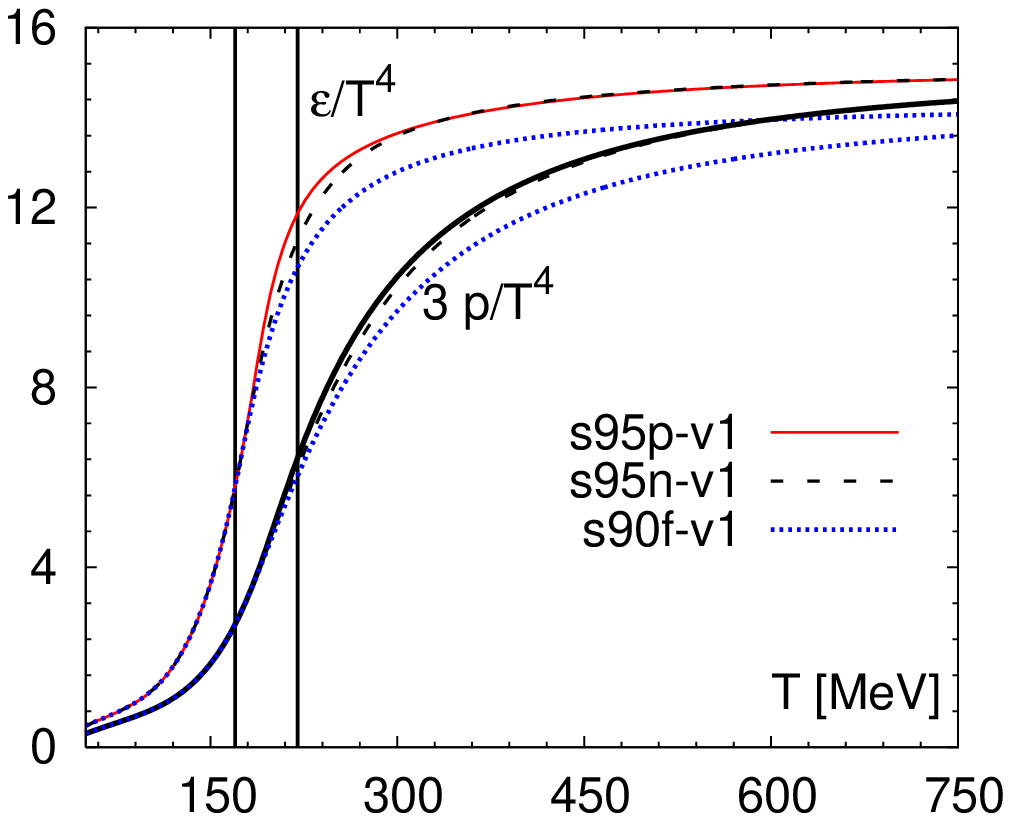}
\includegraphics[width=8cm]{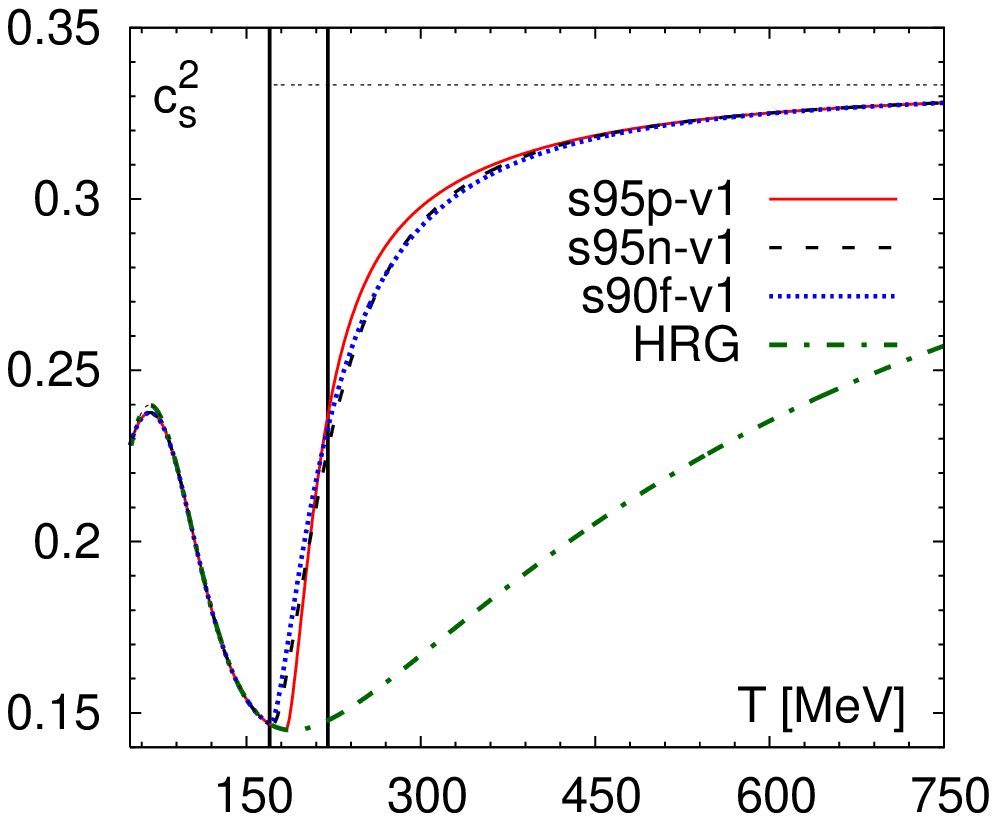}
\caption{The pressure, energy density (left panel) and speed of sound
  (right panel) in the equations of state obtained from
  Eqs. (\ref{e-3p_high}) and (\ref{e-3p_low}). 
The vertical lines indicate the transition region (see text).
In the right panel we also show the speed
of sound for the HRG EoS and EoS with first order phase transition (thin dotted) line, the EoS Q}
\label{fig:EoS}
\end{figure}

The EoSs, obtained by integrating the parametrizations given in
Eqs.~(\ref{e-3p_high}) and (\ref{e-3p_low}) over temperature as shown
in Eqs.~(\ref{P-integral}), are shown in Fig.~\ref{fig:EoS}. The
clearest difference between our different parametrizations is trivial:
The different behavior at high temperatures is due to the different
entropy constraint at $T=800$MeV, which is of course manifested in
energy density and pressure too. On the other hand, the different
height of the peak of the trace anomaly causes only a tiny difference
in pressure and energy density around $T=200$ MeV. This difference is
manifested much clearer in the speed of sound: The large peak and
larger switching temperature from HRG to lattice causes much more
rapid change in the speed of sound in $s95p{\rm -v1}$ parametrization
than in the two others. One could claim that the
differences in the speed of sound are due to the different matching
temperatures $T_0$, but we want to remind the reader that we treated
$T_0$ as a fitting parameter in our model, and any changes in $T_0$
would make the fit to the lattice data worse (see details in
Appendix~\ref{sec:fit}). 
In  Figure \ref{fig:EoS} we show the
speed of sound for EoS\,Q from Refs.~\cite{Kolb,pasi07}: An
equation of state with a first order phase
transition at $T_c=170$MeV. 
Below $T_c$ EoS\,Q coincides with HRG EoS, while above this temperature it is
given by bag equation of state with three massless flavors and bag constant
$B=(0.2447GeV)^4$.
Nevertheless, the most striking feature of the speed of sound in the
proposed parametrization of the 
EoS is that there is no softening below the hadron gas. There is no
region where speed of sound would be smaller than in hadron gas, and
its minimum value is that of HRG speed of sound\footnote{Similar EoS
  was presented already in Refs.~\cite{chojnacki07,chojnacki08}.}.  It
is quite simple to understand why this happens: To achieve smaller
speed of sound than the speed of sound in hadron gas, the trace
anomaly should be larger than in HRG. As one can see in
Fig.~\ref{fig:e-3p}, the present lattice data clearly disfavors such
a scenario.
In Figure \ref{fig:EoS} we indicate the transition region from
hadronic matter to deconfined state by vertical lines. We define the
transition region as the temperature interval $170{\rm MeV} < T < 220
{\rm MeV}$.  In view of the crossover nature of the finite temperature
QCD transition such definition is ambiguous. The lower temperature
limit in our definition comes form the fact that for $T<170$MeV all
our EoS agree with HRG EoS. The rapid rise in the energy density and
entropy density stops roughly at $220$MeV and starting from this
temperature the variation of thermodynamic quantities is quite smooth
\cite{hot_eos}. Also the Kurtosis of the baryon number becomes
compatible with quark gas at temperatures of about $220$MeV
\cite{fluctuations}. For these reasons 
we have chosen it as the upper limit for the transition region.

\subsection{Comparison with other works}
The idea of using HRG in low temperatures and parametrized lattice EoS
in high temperatures is by no means new. Laine and Schr\"oder constructed
QCD equation of state based on the effective theory approach in the high
temperature region, while using the resonance gas equation of state in the low
temperature region \cite{LS}. In the effective theory approach the contribution of hard
modes is treated perturbatively, while the contribution of soft modes is calculated using
3 dimensional lattice simulations of the effective theory called EQCD \cite{eqcd}.
This parametrization has been used in recent recent viscous hydrodynamic calculations \cite{luzum}.
The smooth matching to the resonance gas was done in the temperature interval $T=170-180$MeV.

Two other parametrizations of the
EoS~\cite{chojnacki07,chojnacki08,heinz08} used lattice data of
Budapest-Wuppertal(BW) group obtained using so-called stout staggered
fermion action \cite{BW} and temporal extent $N_{\tau}=4$ and $6$.
Since the stout action does not use higher order difference scheme in
the lattice Dirac operator, discretization effects at high
temperatures are very large. In the ideal gas limit the pressure
calculated on $N_{\tau}=4$ and $6$ lattices is about twice the
continuum value. As the consequence the lattice results of the BW
group have large cut-off effects and overshoot the continuum ideal gas
at high temperatures, see discussion in Ref.~\cite{petr_sewm06}. In an
attempt to correct this problem the authors of Ref.~\cite{BW} divided
all their thermodynamic observables by the corresponding ideal gas
value calculated on $N_{\tau}=4$ and $6$. Since cutoff effects are
strongly temperature dependent this procedure overestimates cutoff
effects in the interesting temperature region and underestimates the
pressure and other thermodynamic observables. 

The Krakow group used the stout lattice results parametrized in
Ref.~\cite{bz} with $T_c=167$MeV and matched the speed
  of sound to the HRG result at that
temperature~\cite{chojnacki07,chojnacki08}.  
The procedure of the Krakow group involves extracting all the
  other thermodynamical quantities from the speed of sound using the
  relation
\begin{equation}
 s(T) = s(T_0)\exp\left[\int_{T_0}^T\frac{dT'}{T'c_s^2}\right].
\label{krakow}
\end{equation}
Connecting the speeds of sound in HRG and lattice leads in this
procedure to a larger entropy density at high temperatures than given
by the lattice parametrization of Ref.~\cite{bz}. To make their EoS to
fit the lattice data at $T\approx 1$GeV, the Krakow group made the
speed of sound smaller in the temperature  
region $28 < T < 118$ MeV
by hand~\cite{chojnacki}. This change is below the expected freeze-out
temperature, and thus the speed of sound which affects the actual
hydrodynamical evolution is unchanged. However, since entropy density
is calculated by integrating over the entire temperature range,
entropy density is smaller than the original HRG
value everywhere in the range $28\mathrm{MeV} < T < T_c$. More
specifically, in the range $120 < T < 160$MeV, both energy and entropy
densities and pressure are $\sim 5\%$ below the HRG value, and energy is
not automatically conserved at freeze-out, see the discussion in
section~\ref{sec:flow}. 

Heinz and Song also used the BW lattice results, but they parametrized
pressure and temperature as function of energy density and matched
them to HRG result; they used $T_c=172$MeV \cite{heinz08}.  As the
authors themselves note, their EoS is not exactly thermodynamically
consistent, which leads to a violation of entropy conservation in an
ideal fluid calculation. However, to our understanding this does not
affect the qualitative studies the EoS has been used so far and the
conclusions of those papers should be valid.

Here a note concerning $T_c$ in these parametrizations is in
order: In Ref.~\cite{BW} thermodynamic quantities are given as
function of $T/T_c$ but the value of $T_c$ is not specified.
At lattice spacings corresponding to temporal extent $N_{\tau}=4$
and $N_{\tau}=6$, used in BW EoS calculations, $T_c$ has large
cutoff effects and may deviate considerably from the continuum
value $T_c=170(3)(4)$MeV determined in Ref. \cite{BW_Tc_09} (e.g.
calculations of $T_c$ reported at Quark Matter 2005 on $N_{\tau}=4$
and $6$ lattices gave $T_c=189(8)$MeV \cite{katz_qm05}). The best
way to eliminate part of the cutoff effects in the BW EoS is to
use the continuum value of the transition temperature $T_c$, which is
interestingly enough appears to agree within errors with the values used in the
phenomenological parametrizations discussed above.

The parametrization of the EoS by the HotQCD
Collaboration~\cite{hot_eos} is based on a simple fit of the lattice
results on the trace anomaly. For temperatures below $130$MeV, where
no lattice data is available, HRG values for the trace anomaly have
been used and assigned an artificial error. The resulting
parametrization is well below the HRG at temperatures $T > 50$ MeV,
see discussion in Ref.~\cite{hot_eos}. For example, at $T\approx
130$MeV and $T \approx 170$MeV temperatures, pressure, energy and
entropy densities are roughly 20\% and 10\% smaller than in HRG,
respectively.  Therefore, when one uses this parametrization, energy
conservation at freeze-out requires additional consideration, see the
discussion in section~\ref{sec:flow}.

\begin{figure}
\includegraphics[width=7.5cm]{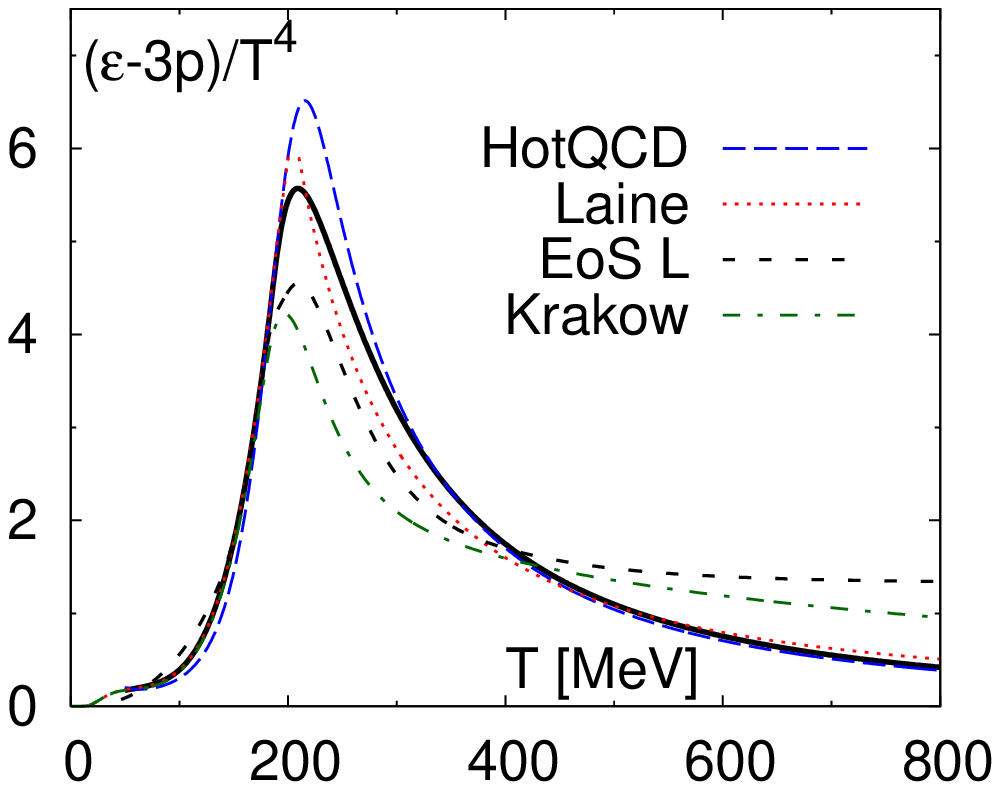}
\includegraphics[width=7.5cm]{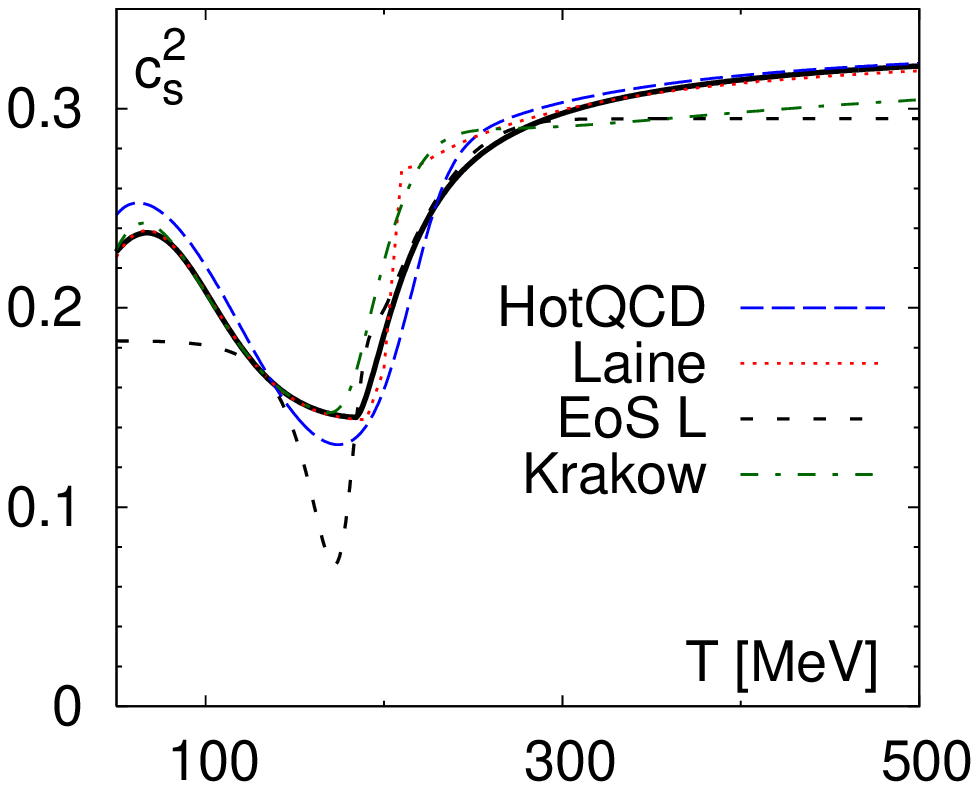}
\includegraphics[width=7.5cm]{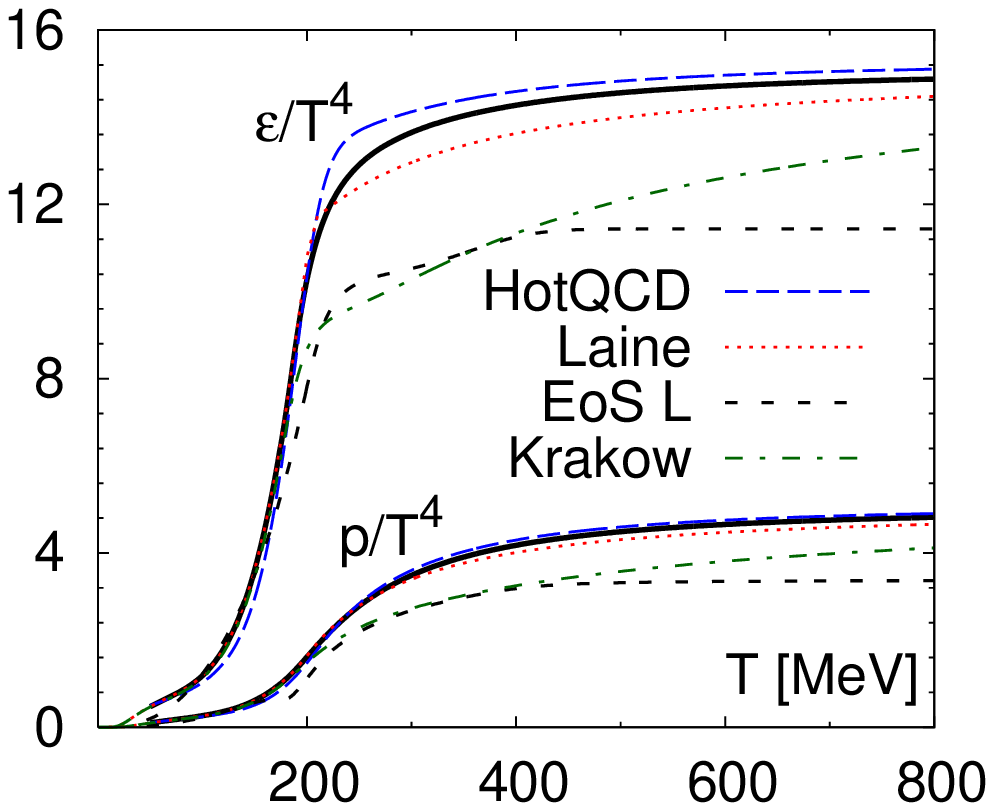}
\includegraphics[width=7.5cm]{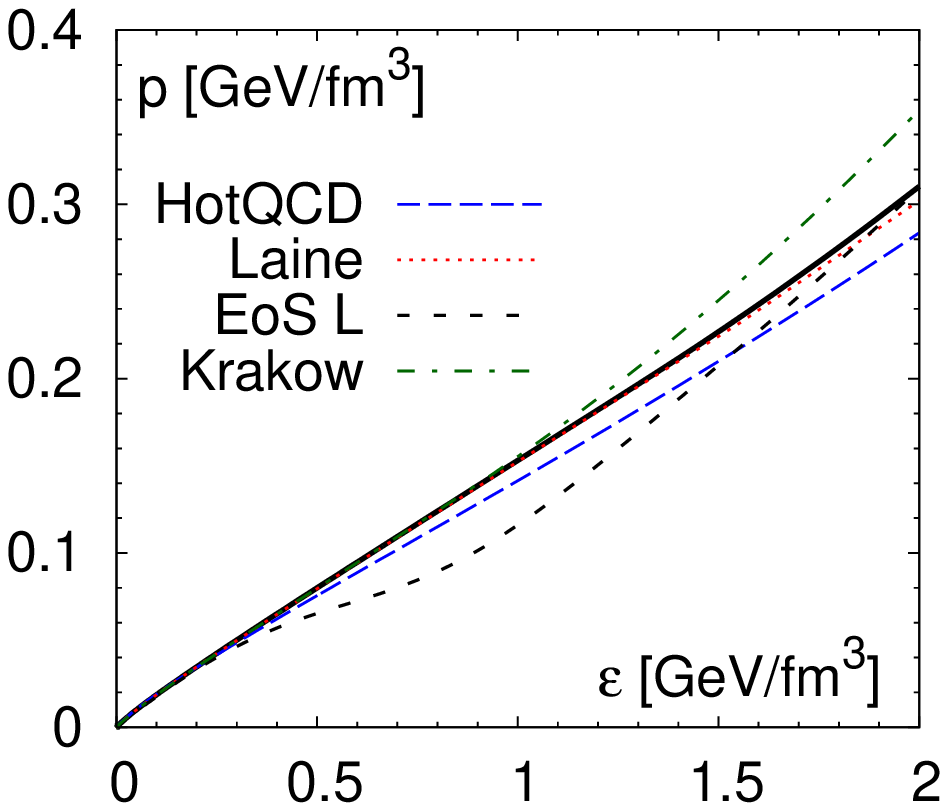}
\caption{The trace anomaly, the speed of sound, the pressure and the energy density as function
of the temperature for different parametrizations used in hydrodynamic models. Also shown is the
the pressure as function of the energy density. The solid black line corresponds to {\it s95p}-v1
parametrization.}
\label{fig:comp}
\end{figure}

In the Fig.~\ref{fig:comp} we show the comparison of our
parametrization for EoS with the ones discussed above including the
trace anomaly, speed of sound, pressure and energy density as function
of the temperature as well as the pressure as function of the energy
density.
From the figure we see that there are significant differences between different 
parametrizations. The parametrization based on BW lattice results seem to have different
high temperature asymptotic. In the low temperature region the EoS~L parametrization by 
Heinz and Song and the HotQCD parametrization differ significantly
from others. It is worth noting that EoS\,L does not even try to
  reproduce the HRG  below 130 MeV temperature. The authors assumed
  that the details of the EoS below freeze-out temperature have only a
  negligible effect on the evolution within the freeze-out surface and a faithful reproduction of the HRG is thus not needed.
Finally, for the trace anomaly, the difference between our parametrization and HotQCD
parametrization is limited to temperatures $T<250$MeV by
construction. However, since the pressure, energy density and entropy density is
obtained using the integral method the differences in trace anomaly at low temperatures
result in differences at all temperatures for these quantities.

\section{Effect on hydrodynamical flow}
  \label{sec:flow}
Now we are in a position to use the lattice results on EoS in a hydrodynamical model
and compare the results with previous approaches using first order phase transition or the other lattice parametrizations 
in the literature. We also quantify how present uncertainties in the EoS parametrization affect 
the hydrodynamical flow. For simplicity we perform our analysis using ideal hydrodynamics. 
This is also motivated by the fact that the overall value and temperature dependence of the QCD 
and HRG transport 
coefficients is not known and any attempts to parametrize them would introduce additional uncertainties 
in the analysis. 

\begin{figure}
\includegraphics[width=8.0cm]{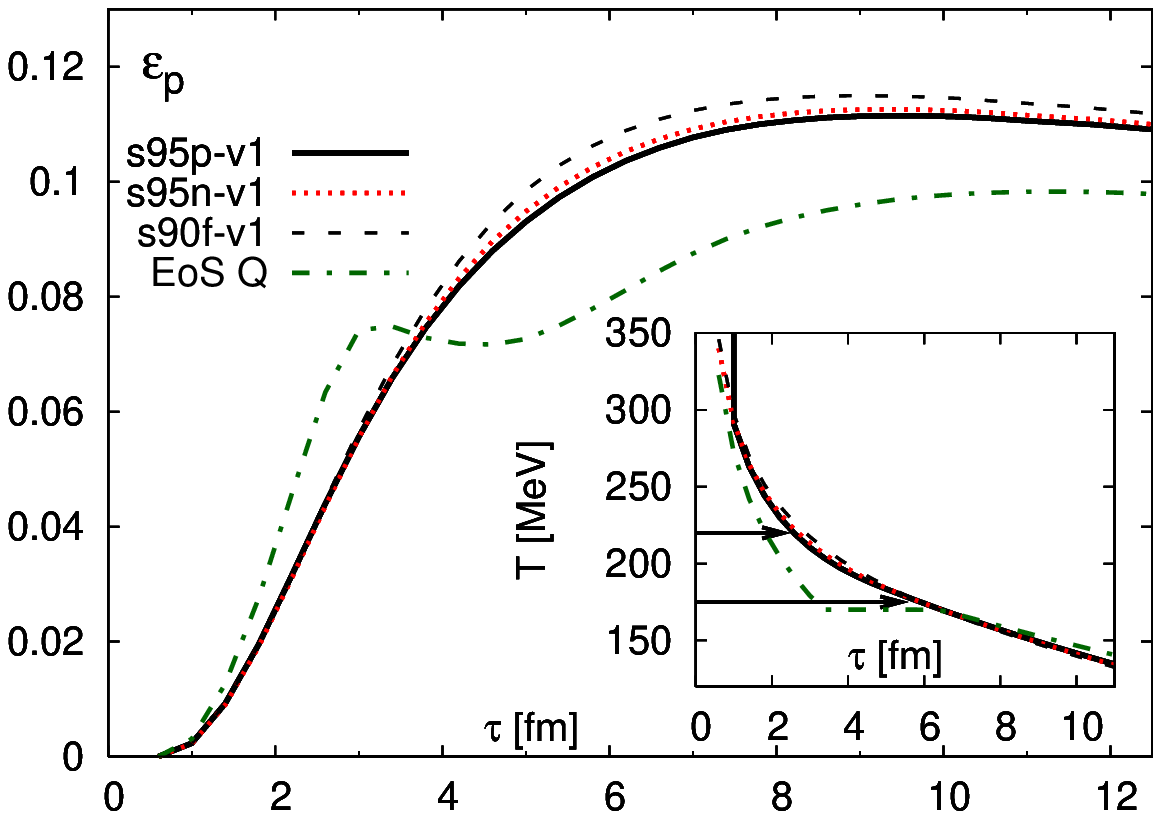}
\includegraphics[width=8.0cm]{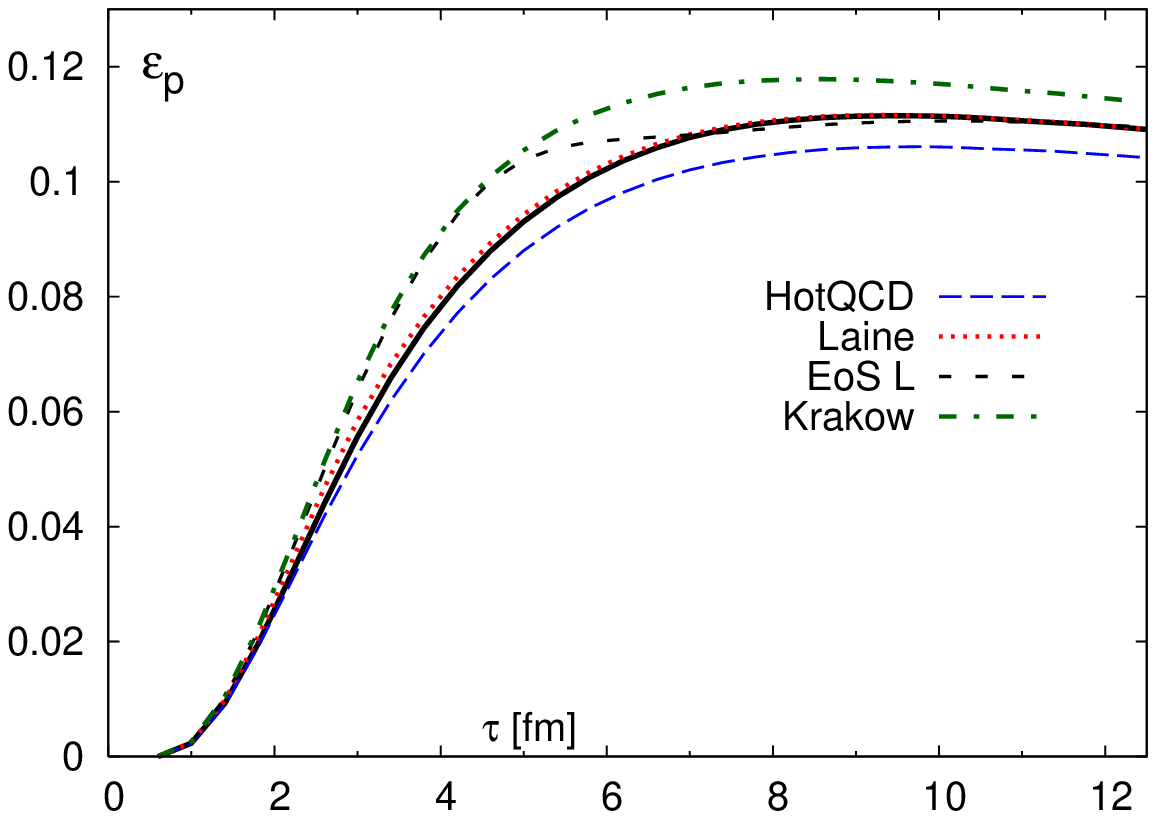}
\caption{The time evolution of the momentum anisotropy in $b=7$ fm
  Au+Au collisions using the EoSs developed in this paper and the old
  EoS with a first order phase transition (EoS\,Q) from
  Ref.~\cite{Kolb,pasi07} (left panel) and the EoS $s95p\mathrm{-v1}$
  compared to the various lattice EoSs in the
  literature~\cite{LS,chojnacki07,chojnacki08,heinz08,hot_eos} (right
  panel). In the right panel the solid black line refers to the result obtained
  with $s95p\mathrm{-v1}$ parametrization. 
  The inset in the right panel shows the temperature
  evolution in the middle of the system for different EoS. The
  horizontal lines indicate the transition region.}
\label{fig:epsilon_p}
\end{figure}

As the first step we study the sensitivity of the momentum anisotropy $\epsilon_p$ on
the EoS. This is the cleanest way to address the sensitivity of hydrodynamic flow
to the EoS as additional complications due to 
freezeout do not enter here.
The momentum anisotropy is defined as~\cite{Kolb} 
\begin{equation}
\epsilon_p=\frac{\langle T_{xx}-T_{yy}\rangle}{\langle T_{xx}+T_{yy}\rangle}, 
\end{equation}
where $T_{xx}$ and $T_{yy}$ are the diagonal transverse components of
the energy-momentum tensor and brackets denote averaging over the
entire transverse plane. In Figure \ref{fig:epsilon_p} we show the
time evolution of momentum anisotropy in Au+Au collisions at
$\sqrt{s_\mathrm{NN}}=200$ GeV with $b=7$fm impact parameter for
different EoSs.  The left panel shows the anisotropy calculated using
our different parametrizations, and EoS~Q from
Refs.~\cite{Kolb,pasi07}. 
As studied in detail in
Ref.~\cite{Kolb} the first order phase transition causes the build up
of the flow anisotropy to stall when most of the system is in the
mixed phase. There is no such a structure when the transition to the
hadronic matter is a smooth cross over, but the anisotropy increases
monotonously. The hardness of the EoS in plasma phase is also
manifested in the early behavior of the anisotropy. EoS\,Q is much
harder in that region than any of the lattice EoSs studied here, and
thus the build up of the flow anisotropy is faster. On the other hand,
the mixed phase makes the EoS\,Q much softer in average during the
evolution, and the final anisotropy is the smallest of all EoSs
studied here. The speed of sound is quite similar in EoSs 
$s95p\mathrm{-v1}$, $s95n\mathrm{-v1}$ and $s90f\mathrm{-v1}$, and consequently the
development of the flow anisotropy is similar.  When the 
system cools, the speed of sound stays large longest for $s95p\mathrm{-v1}$, but it also 
drops fastest and stays small longest for
$s95p\mathrm{-v1}$. These effects cancel, and the evolution of the
anisotropy is almost identical to $s95n\mathrm{-v1}$.  After that
argument, the largest anisotropy obtained using $s90f\mathrm{-v1}$ may
look surprising, but closer inspection of Fig.~\ref{fig:EoS} reveals
that $s90f\mathrm{-v1}$ has always either larger or equal speed of sound
than $s95p\mathrm{-v1}$. Thus $s90f\mathrm{-v1}$ is harder, and it 
should lead to a larger anisotropy than $s95p\mathrm{-v1}$.

The old wisdom has been that elliptic flow builds up quickly during
the early stages of the evolution and is mostly build up during the
plasma phase. For example, for EoS\,Q, three quarters of the final
anisotropy has been built up when the center of the system reaches
mixed phase. For lattice based EoSs this is no longer as clear:
Roughly half of the anisotropy is built up during the transition
region, after the first three fm of the evolution, but the
hadronic contribution from below $T=170$ MeV temperatures to
$\epsilon_p$ is essentially negligible unlike in
EoS Q. This difference in the evolution is partly explained by
the softer EoS in the plasma phase---anisotropy is built up
slower. Another reason is that the transition region in our EoSs
reaches up to $\epsilon \approx 3.5$ GeV/fm$^3$ energy density,
whereas EoS\,Q reaches plasma phase already at $\epsilon\approx 2.15$
GeV/fm$^3$ density.

In the right panel of Fig.~\ref{fig:epsilon_p} $s95p-\rm v1$ is compared to
other lattice EoSs in the literature. The differences in flow are
difficult to sort out based on the speed of sound as function of
temperature alone, but they are easy to understand when one looks at
the pressure as function of energy density (Fig.~\ref{fig:comp}).
It can be seen that the gradient $\partial P/\partial \epsilon$,
i.e.~speed of sound squared, is largest for the Krakow EoS, and EoS L
is approximately as stiff as the Krakow EoS above $\epsilon \approx 1$
GeV/fm$^3$ density. Thus it is not surprising that the Krakow EoS
leads to the largest anisotropy, and that the initial build up of the
anisotropy is similar for the Krakow EoS and EoS L. The build up of
flow deviates when EoS L reaches its softest region which is much
softer than in any other EoS, and leads to behavior reminiscent of
EoS\,Q: A sudden stall in the increase of the anisotropy. Unlike in EoS
Q, however, there is no subsequent increase in anisotropy after the
soft region has been passed. Likewise, even if the speed of sound in
the HotQCD EoS at high temperatures is equal or even larger than in
$s95p\mathrm{-v1}$ or in the EoS by Laine and Schroder, the speed of sound in the
transition region is so much smaller than in the other EoSs, that the
final anisotropy is the smallest of all. As well, even if the speed of
sound as function of temperature look different for $s95p\mathrm{-v1}$ and Laine
and Schroder's EoS, as a function of energy density they are almost
equal, and thus these EoSs lead to basically identical build up of the
flow anisotropy.

\begin{figure}
\includegraphics[width=6.0cm]{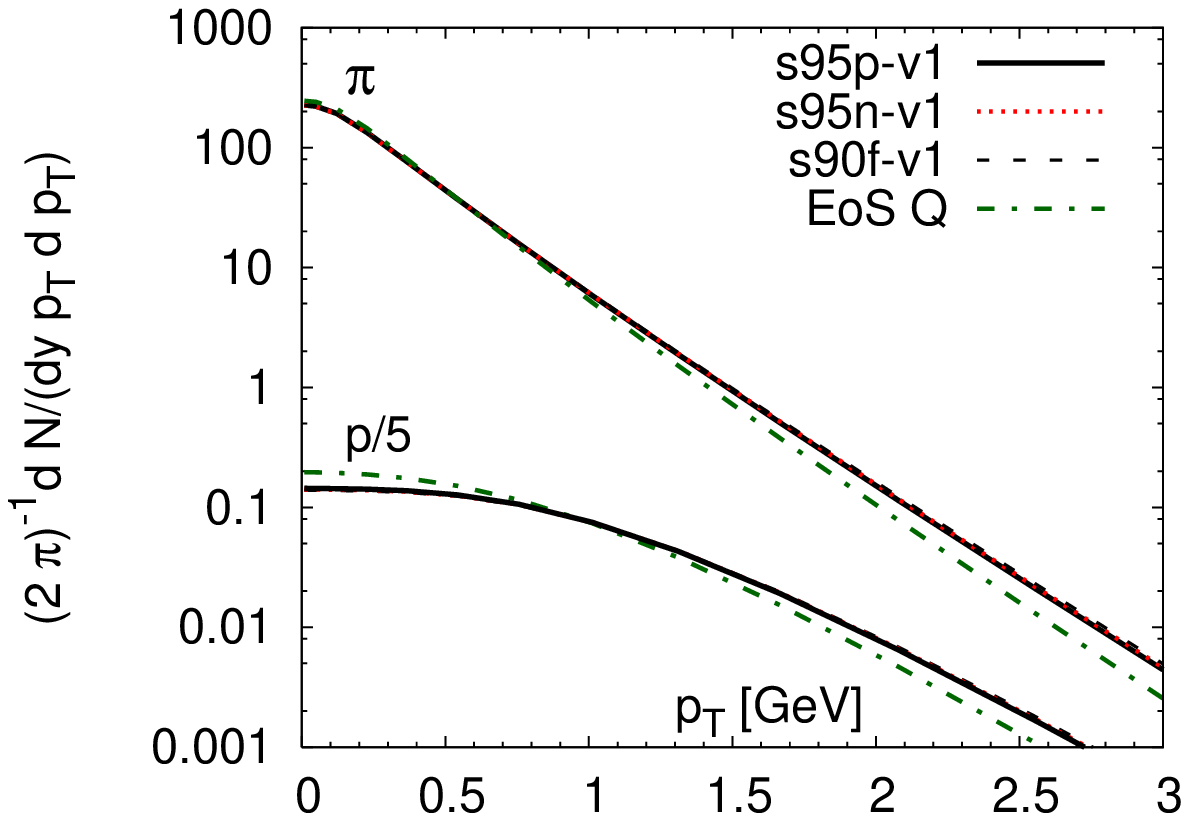}
\includegraphics[width=5.0cm]{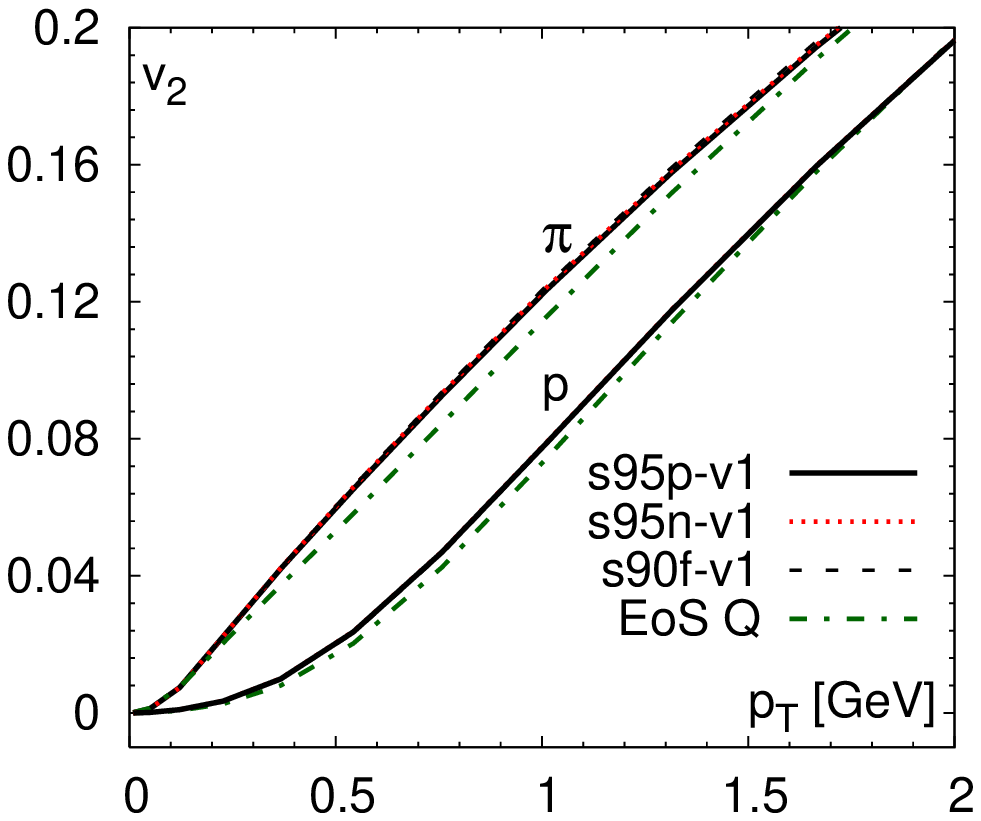}
\includegraphics[width=5.0cm]{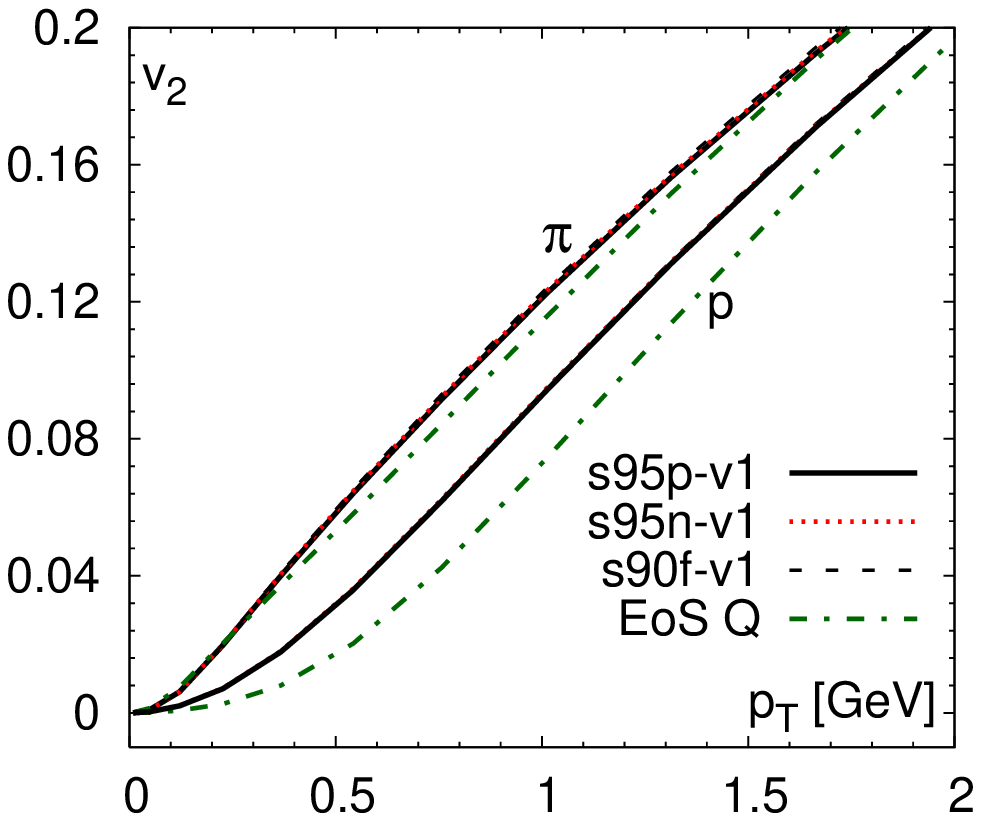}
\caption{The proton and pion spectra (left) and differential elliptic
  flow $v_2(p_T)$ of protons and pions (middle and right) 
  in $b=7$fm Au+Au collisions for
  different EoSs. The results in the left and middle panels are
  calculated using the same freeze-out temperature $T_\mathrm{fo}=125
  MeV$ for all the EoSs, whereas in the right panel it has been
  adjusted to produce similar $p_T$-distributions. $T_\mathrm{fo}=125
  MeV$ for EoS\,Q, and $T_\mathrm{fo}=140 MeV$ for EoSs $s95p\mathrm{-v1}$,
  $s95n\mathrm{-v1}$ and $s90f\mathrm{-v1}$.}
\label{fig:flow_fixed} 
\end{figure}

As the next step we study the sensitivity of the spectra and elliptic
flow on the EoS.  We include freeze out into the calculation described
above, and use first the same freeze out temperature,
$T_\mathrm{fo}=125$MeV, for all EoSs\footnote{This temperature was
  found to reproduce the spectra when EoS\,Q is used~\cite{pasi05}.}.
The pion and proton spectra after resonance decays is shown in the
left panel of Fig.~\ref{fig:flow_fixed} for EoSs $s95p\mathrm{-v1}$, 
$s95n\mathrm{-v1}$, $s90f\mathrm{-v1}$ and EoS\,Q. As expected the new
parametrization lead to flatter spectra than EoS\,Q, but the
differences between the parametrizations themselves are too small to
result in significant differences in spectra. The $p_T$-differential
$v_2$ of pions and protons shown in the middle panel are surprisingly
insensitive to the EoS. The larger flow anisotropy shown in
Fig.~\ref{fig:epsilon_p} leads to larger $p_T$ averaged $v_2$, but
that is mostly due to flatter spectra weighting $v_2(p_T)$ at higher
$p_T$ where it is larger, than due to $v_2(p_T)$ being larger. One
must also remember that this result is obtained using the same
freeze-out temperature for all the EoSs. Before discussing how the EoS
affects elliptic flow, one has to readjust the freeze-out temperature
to produce similar spectra. This we have done in the right panel of
Fig.~\ref{fig:flow_fixed}: When one uses $T_\mathrm{fo}=140$MeV for
EoSs  $s95p\mathrm{-v1}$, $s95n\mathrm{-v1}$ and $s90f\mathrm{-v1}$, 
the spectra are similar to those calculated using EoS\,Q. The pion $v_2(p_T)$ is
virtually insensitive to the change in freeze-out temperature, but the
higher temperature leads to much larger $p_T$-differential $v_2$ for
protons. This behavior has already been explained in
Ref.~\cite{Huovinen:2001}, where it was argued that the lower the
temperature and larger the flow velocity, the smaller the $v_2(p_T)$
at low values $p_T$, and that the heavier the particle, the stronger
this effect. Note that the three different lattice based
parametrizations of EoS give almost identical results for $v_2$ and
spectra. This means that existing uncertainties in the EoS
parametrization have negligible effect on the flow.

Unfortunately it is not straightforward to calculate the spectra and
$v_2$ using the various EoSs in the literature discussed earlier. One
of the advantages of the Cooper-Frye procedure for freeze-out is that
energy, momentum, particle number and entropy are conserved. But, they
are conserved only if the equation of state is the same before and
after the freeze-out~\cite{Laszlo}, i.e.~that the fluid EoS is that of
free particles and that the number of degrees of freedom in the fluid
is the same than the number of hadrons and resonances which spectrum
is calculated. This is not the case with the EoSs discussed here. When
calculating the spectra we use the same set of resonances up to 2 GeV
mass than what is included in our HRG. Laine and Schroder used
resonances up to 2.5 GeV mass, but at, say, $T_\mathrm{fo}=125$ MeV
freeze-out temperature the difference in energy and entropy is
minuscule, about 0.05\%. The situation with the other EoSs is more
difficult. At the mentioned temperature, the Krakow EoS has 4.5\%
smaller, EoS\,L 7\% larger, and the HotQCD EoS 22\% smaller energy and
entropy than hadrons and resonances up to 2 GeV mass. One way to
correct this discrepancy is of course to change the number of hadrons
and resonances included in the spectra calculation. But, this approach
would be tedious since the number of resonances needed to fit the
densities in the EoS may depend on temperature. Also there is no
telling whether there exists a set of resonances reproducing both
densities and pressure at a given temperature, and the number of
resonances could be surprisingly small. For example a hadron resonance
gas consisting of only pseudo-scalar and vector meson nonets, and
baryon octet and decuplet, still has $~10$\% larger energy density at
$T=125$ MeV than HotQCD EoS. Therefore we follow the approach espoused
by Csernai~\cite{Laszlo} and Bugaev~\cite{Bugaev}: We require that
energy and momentum are conserved locally on the freeze-out surface, i.e.
\begin{equation}
  \mathrm{d}\sigma_\mu T_\mathrm{fluid}^{\mu\nu}
  = \mathrm{d}\sigma_\mu T_\mathrm{particles}^{\mu\nu},
\end{equation} 
where $T_\mathrm{fluid}^{\mu\nu}$ is the energy-momentum tensor of the
fluid on the surface, and $T_\mathrm{particles}^{\mu\nu}$ is the
energy-momentum tensor of the emitted particles. To conserve energy
and momentum, we allow the temperature and flow velocity of fluid and
particles \emph{differ}, i.e. there is a discontinuity on the
  surface, and freeze-out is a shock-like
phenomenon~\cite{Bugaev}. We have to admit the corrections due to this
procedure are small and mostly affect the multiplicity, but consider
obeying the conservation laws worth the extra effort.

At first we use the same freeze-out energy density
we used when comparing our parametrization to
EoS\,Q, $\epsilon = 0.065$ GeV/fm$^3$, for all EoSs. The
corresponding temperatures for each EoS are listed in
Table~\ref{tab:Tdec}.
\begin{table}
\begin{tabular}{c|ccc|ccc}
\hline
         & $\epsilon_\mathrm{fo}$ & $T_\mathrm{fluid}$ & $T_\mathrm{particles}$
         & $\epsilon_\mathrm{fo}$ & $T_\mathrm{fluid}$ & $T_\mathrm{particles}$ \\
\hline      
$s95p-\mathrm{v1}$   & 0.065 & 125 & 125 & 0.14  & 140 &  140 \\
HotQCD   & 0.065 & 129 & 125 & 0.11  & 139 &  135 \\
Laine    & 0.065 & 125 & 125 & 0.14  & 140 &  140 \\
EoS\,L   & 0.065 & 124 & 126 & 0.11  & 134 &  136 \\
Krakow   & 0.065 & 126 & 125 & 0.185 & 146 &  145 \\
\hline
\end{tabular}
\caption{The freeze-out energy densities (in GeV/fm$^3$) and corresponding
  temperatures (in MeV) for fluid and particles for each EoS used in the
  calculations shown in Fig.~\ref{fig:flow_market}}
\label{tab:Tdec} 
\end{table} 
\begin{figure}
\includegraphics[width=6.0cm]{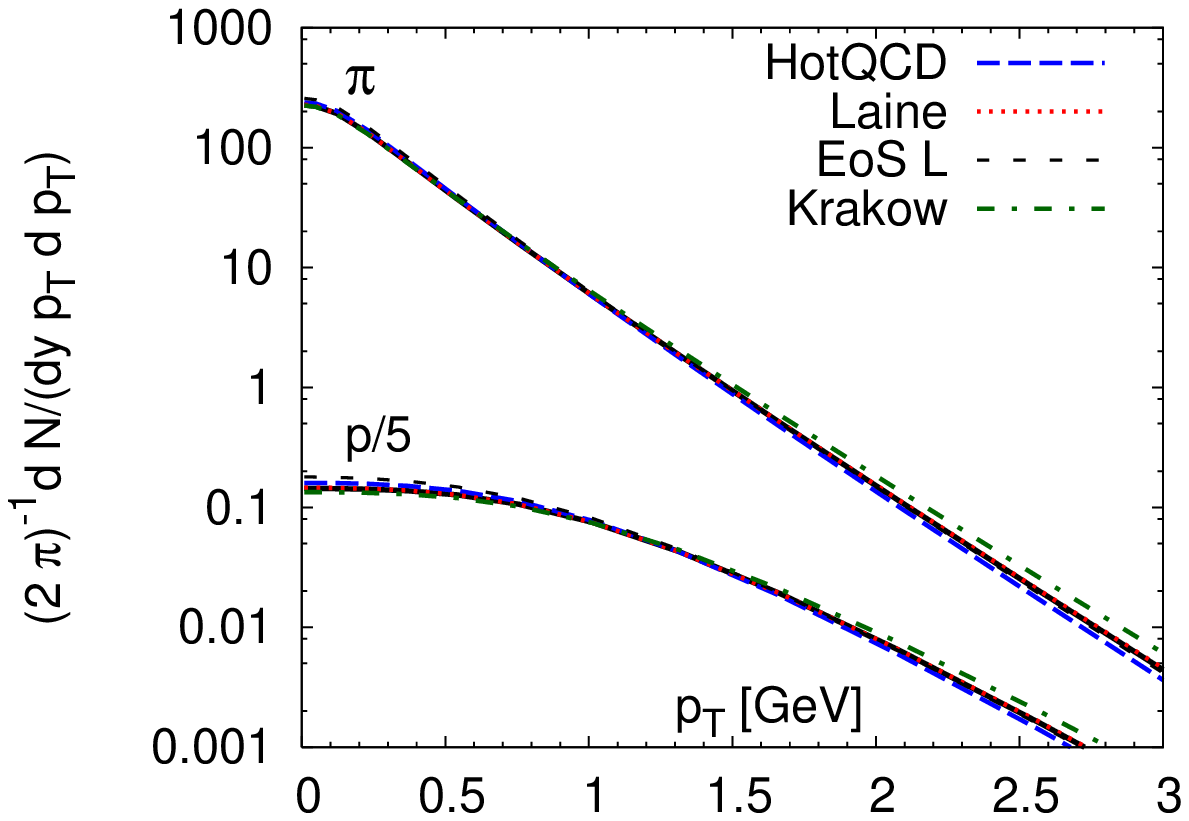}
\includegraphics[width=5.0cm]{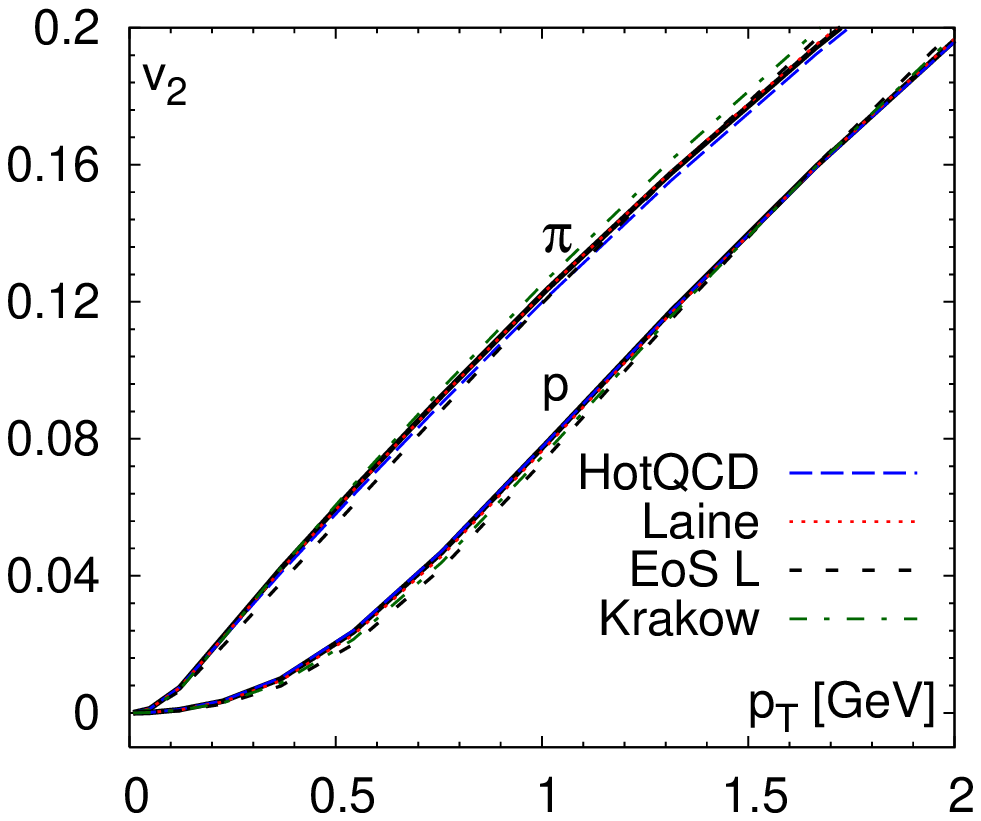}
\includegraphics[width=5.0cm]{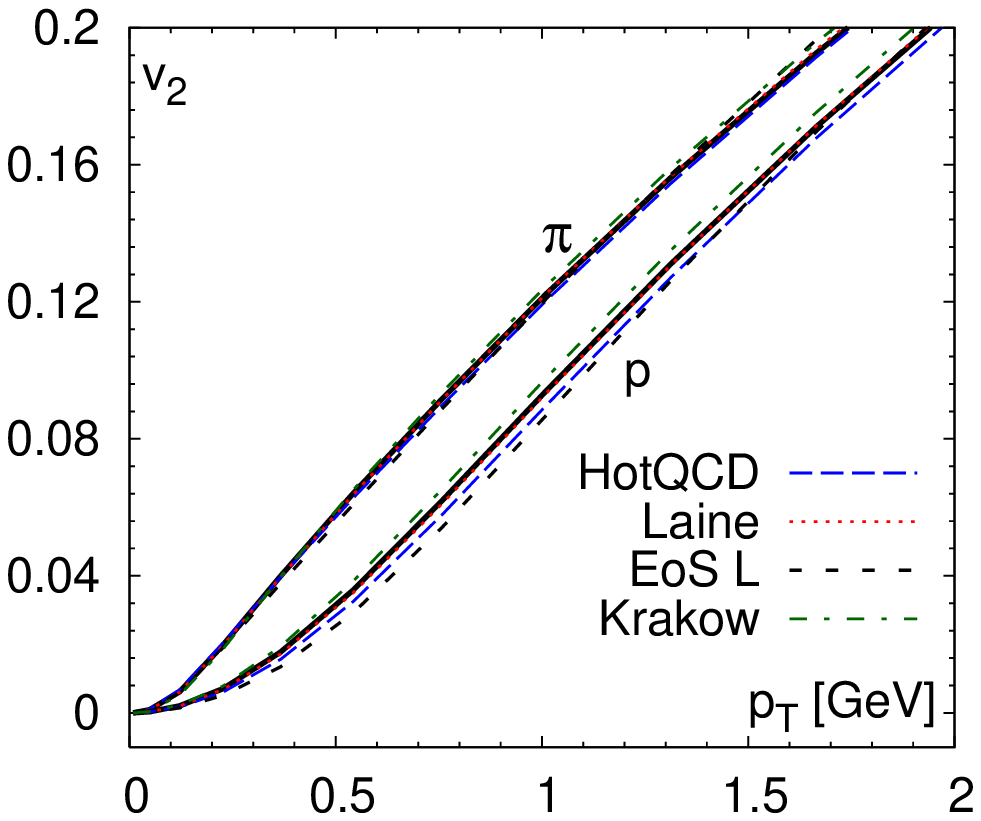}
\caption{The proton and pion spectra (left) and differential elliptic
  flow $v_2(p_T)$ of protons and pions (middle and right) in $b=7$fm
  Au+Au collisions for $s95p\mathrm{-v1}$ (solid black line) and different EoSs in the
  literature~\cite{LS,chojnacki07,chojnacki08,heinz08,hot_eos}.  The
  results in the left and middle panels are calculated using the same
  freeze-out energy density $\epsilon_\mathrm{fo}=0.065$ GeV/fm$^3$
  for all the EoSs, whereas in the right panel it has been adjusted to
  produce similar $p_T$-distributions, see text and
  Table~\ref{tab:Tdec}}
\label{fig:flow_market} 
\end{figure}
In the left panel of Fig.~\ref{fig:flow_market} we show the $p_T$
distributions of pions and protons in $b=7$ fm Au+Au collision.  The
differences in distributions are small, and the general behavior is
what can be expected based on the stiffness of the EoSs and the flow
anisotropy: The Krakow EoS is the stiffest, and leads thus to the
flattest spectra. The EoS by Laine and Schr\"oder leads to behavior
almost identical to $s95p$-v1, and the HotQCD EoS and EoS\,L are the
softest and have slightly steeper spectra than the other EoSs. For the
$p_T$-differential anisotropy of pions and protons the systematics is
the same than seen for our parametrizations: when the freeze-out
criterion is the same for all the EoSs, $v_2(p_T)$ is basically
independent of the EoS, as shown in the middle panel of
Fig.~\ref{fig:flow_market}. After the freeze-out criterion is adjusted
to reproduce spectra obtained using EoS\,Q, the pion $v_2(p_T)$ is
independent of the EoS, but the proton anisotropy shows some
sensitivity, see the right panel of Fig.~\ref{fig:flow_market}. The
differences between the EoSs are small and thus the differences in
$v_2(p_T)$ are small, but an ordering according to the stiffness of
the EoS is visible: The Krakow EoS is hardest, and its proton
$v_2(p_T)$ is largest at small $p_T$, whereas the HotQCD EoS and
EoS\,L are softest and lead to lowest $v_2(p_T)$ of protons at low
$p_T$. After all the main results of this comparison are that the
differences in the lattice EoS parametrization in the literature are
small and not observable in the $p_T$-differential elliptic flow, and
that energy conservation at freeze-out is not trivial if the EoS at
freeze out is not that of free hadron resonance gas.

\begin{figure}
\includegraphics[width=8.7cm]{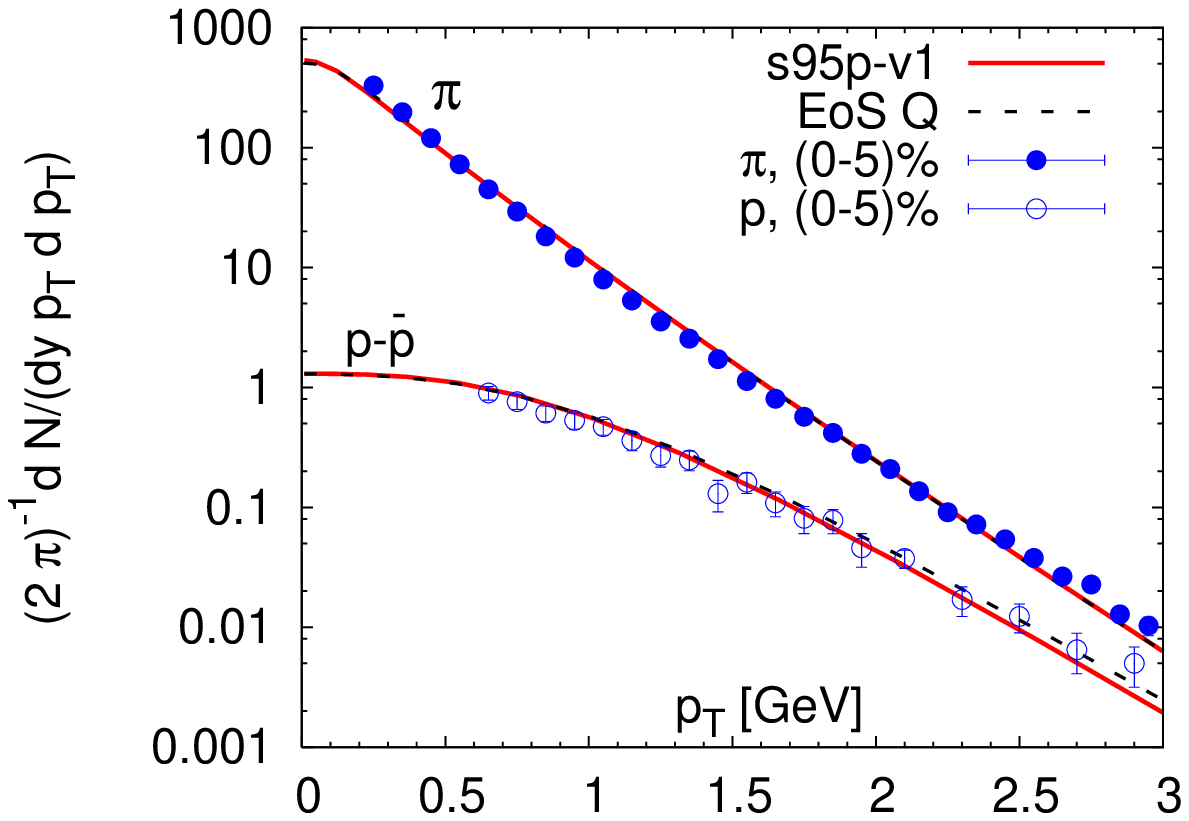}
\includegraphics[width=7.4cm]{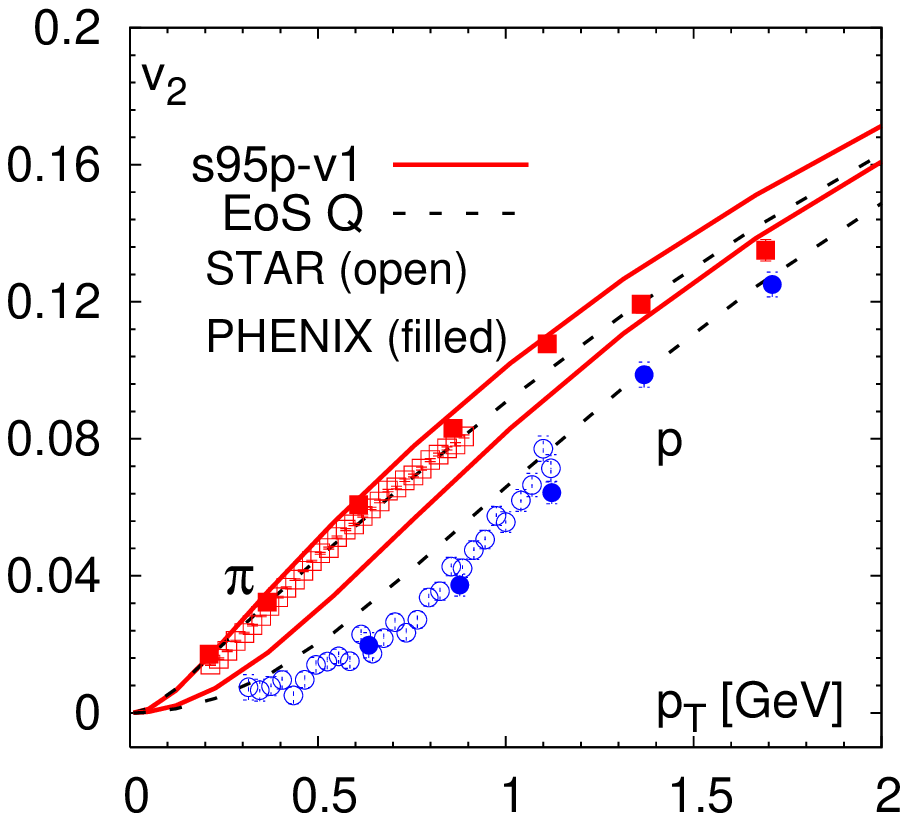}
\caption{Pion ($\pi^+$) and net-proton ($p-\bar{p}$) spectra in 0-5\%
  most central (left), and pion and antiproton $p_T$-differential
  elliptic flow $v_2(p_T)$ in minimum bias (right) Au+Au collisions at
  $\sqrt{s_\mathrm{NN}}=200$ GeV compared with hydrodynamic
  calculations using two different EoSs and assuming chemical
  equilibrium. The data was taken by the PHENIX~\cite{Phenix} and the
  STAR~\cite{Star} collaborations.  }
\label{fig:flow_fit}
\end{figure}

Finally we want to compare the results of our calculations with
data. Since all of our parametrizations lead to practically identical
spectra and $v_2$, we use only $s95p$-v1 for simplicity, and compare
the results to those obtained using EoS\,Q in
Ref.~\cite{pasi05}. First we fix all the parameters by requiring the
reproduction of pion and net-proton ($p-\bar{p}$) spectra in the 0-5\%
most central Au+Au collisions at $\sqrt{s_\mathrm{NN}}=200$ GeV
energy. The resulting spectra are shown in the left panel of
Fig.~\ref{fig:flow_fit}, and the freeze-out temperatures are the same
than mentioned before, $T_\mathrm{fo}=125 MeV$ for EoS\,Q, and
$T_\mathrm{fo}=140 MeV$ for EoSs $s95p-\mathrm{v1}$. Since we assume
chemical equilibrium we cannot reproduce both proton and anti-proton
yields at such a low temperature. Our parametrization is for zero
baryochemical potential, so we cannot calculate net-protons either,
but we approximate them by having a finite baryon density in the
calculation, and converting this density into a finite chemical
potential at freeze-out by using a HRG EoS which allows a finite net
baryon density. The right panel of Fig.~\ref{fig:flow_fit} shows the
$p_T$-differential elliptic flow of pions and anti-protons in minimum
bias Au+Au collisions at the same energy. As earlier, the pion
$v_2(p_T)$ is very similar for both EoSs, but the anti-proton
$v_2(p_T)$ is quite different. In fact for the realistic EoS the
anti-proton $v_2$ is largely overpredicted, while we have a reasonable
agreement with the data when the bag model EoS\,Q is used. This is
very similar to the finding of Ref.~\cite{pasi05}. Also the Krakow
group seem to overpredict the proton $v_2$ \cite{chojnacki08}
indicating that this could be a general feature of ideal hydrodynamic
models using more realistic EoS. The first results indicate that
dissipative effects can at least reduce this
problem~\cite{Bozek:2009}.

This analysis has been repeated assuming partial
chemical equilibrium in the hadronic phase. The details of this
analysis are given in the appendix~\ref{sec:PCE}. Our main finding did
not change. Using different EoS for the same initial conditions and
kinetic freeze-out temperature has little effect on $v_2$ but leads to
significant differences in the spectra. If the spectra are fitted to
reproduce the experimental data by adjusting the initial conditions,
the $p_T$-differential elliptic flow of anti-protons becomes too large for
lattice based EoS, but agrees reasonably well with
EoS\,Q. 
On the other hand, the $p_T$-differential elliptic flow of
pions is clearly too large for both EoSs requiring substantial
dissipation to reduce it to fit the data.

\section{Conclusion}

In this paper we addressed the question to what extent the Hadron Resonance Gas (HRG) model can describe
the thermodynamic quantities calculated on the lattice. As discussed above this question is very important
for implementing a consistent freeze-out prescription or a consistent switch to transport in hydrodynamic or hybrid
models, respectively. We have found that lattice data
strongly disagree with the HRG model in the low temperature regime. The reason for this disagreement has been
identified with large cutoff effects in the lattice calculations. We also showed that taking into account the
discretization effects in the hadron spectrum in the HRG model leads to a good agreement with the lattice data.
In fact, we find that for some quantities the HRG model works well to unexpectedly high temperatures. 
Based on this observation we constructed several parametrizations of the equation of state which interpolate between
the lattice data at high temperature and the resonance gas in the low temperature region. The central quantity
in this analysis was the trace anomaly since it is directly calculated on the lattice and the differences in
the proposed parametrizations are found in the temperature region where the trace anomaly reaches its maximal value.
 
We studied the hydrodynamic evolution using three parametrizations of
the EoS that interpolate between HRG EoS and the lattice data and
compared the results with the corresponding ones obtained using an EoS
with a first order phase transition, the so-called EoS\,Q, as well as
several other parametrizations of the EoS used in the literature.  We
have analyzed the flow in terms of momentum space anisotropy
$\epsilon_p$, $p_T$-differential elliptic flow $v_2(p_T)$ and proton
and pion spectra. The three parametrizations of the EoS proposed in
this paper as well the the parametrization by Laine and Schr\"oder
\cite{LS} gave very similar results for all of the above quantities.
The effect of using different EoS parametrizations is the most visible
in $\epsilon_p$. The difference in the results obtained with EoS\,Q and
other parametrizations is especially large.  Quite surprisingly
$v_2(p_T)$ is not sensitive to the choice of the EoS if the same 
freeze-out temperature is used. The particle spectra
on the other hand are sensitive to the EoS. However, the change in the
EoS can be compensated by change of the freeze-out temperature. If the
freeze-out temperature is adjusted to reproduce the particle spectra we
see large differences in the proton $v_2(p_T)$ for EoS\,Q and other EoS
parametrizations. However, for all the other parametrizations considered
here, the proton $v_2(p_T)$ is quite similar.

The work presented in this paper should be extended in number of different ways. First we should extend the
comparison of lattice QCD results with modified HRG to other fluctuations, including electric charge fluctuations
as well as fourth and higher order fluctuations of baryon number, strangeness and electric charge. 
However, since these quantities were studied in detail only for the p4 action, in the analysis additional assumptions about
the cutoff effects in the hadron spectrum have to be made. Furthermore, it would be interesting to study the quark
mass effects in the HRG model, especially since recent lattice calculations of the EoS extend to physical
values of the light quark masses \cite{p4_new_eos}. 
We also should
consider the effect of finite baryon potential on the EoS. This will become important for application of hydrodynamic
models to heavy ion collisions at lower energies, especially to the proposed RHIC energy scan. Finally, it will be interesting
to study the effect of the EoS in the framework of viscous hydrodynamics. 
We plan to address these issues in forthcoming
publications.

\section*{Acknowledgement}
This work was supported by the U.S. Department of Energy under
contract DE-AC02-98CH1086 and by the ExtreMe Matter Institute
(EMMI). P.H.~is grateful for support from Center of Analysis and
Theory for Heavy Ion Experiment (CATHIE) which enabled him to stay in
BNL where large part of this work was finalized, and for hospitality
for Iowa State University where part of this work was done. 
We thank Larry McLerran for encouraging us to do this work.
We also
want to thank Mikko Laine, Mikolaj Chojnacki, Wojciech Florkowski,
Huichao Song and Ulrich Heinz for providing us with their equations of
state for comparison.
P.P. thanks the members of the MILC collaborations, especially Carleton DeTar, Steve Gottlieb, Urs Heller and Bob Sugar
for correspondence and for providing their numerical results, including the unpublished data of Ref.~\cite{milc_unpub}.
P.P. is also grateful to Zolt\'an ~Fodor and S\'andor Katz for correspondence 
and for providing the pion masses for the stout action.

\appendix

\section{Hadron masses on the lattice}

\begin{figure}
\includegraphics[width=10cm]{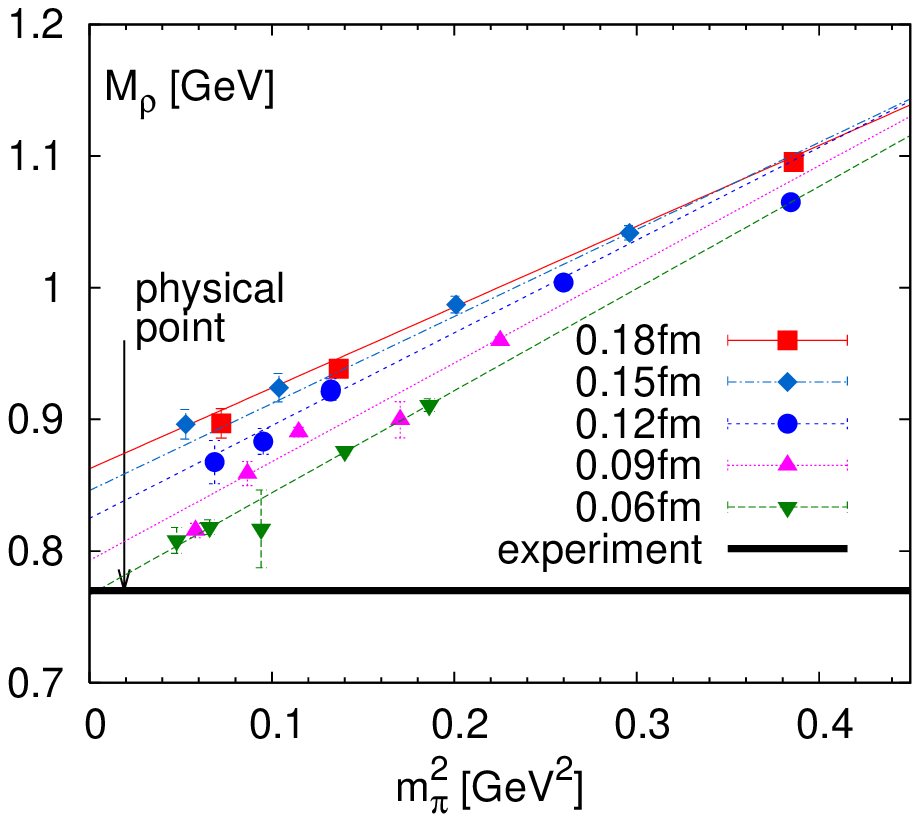}
\caption{The $\rho$-meson (right) masses calculated on the lattice using
asqtad action \cite{bernard01,bernard04,milc_unpub} and compared with Eq. (\ref{fit}) (lines).} 
\label{fig:mrho}
\end{figure}

\begin{figure}
\includegraphics[width=8cm]{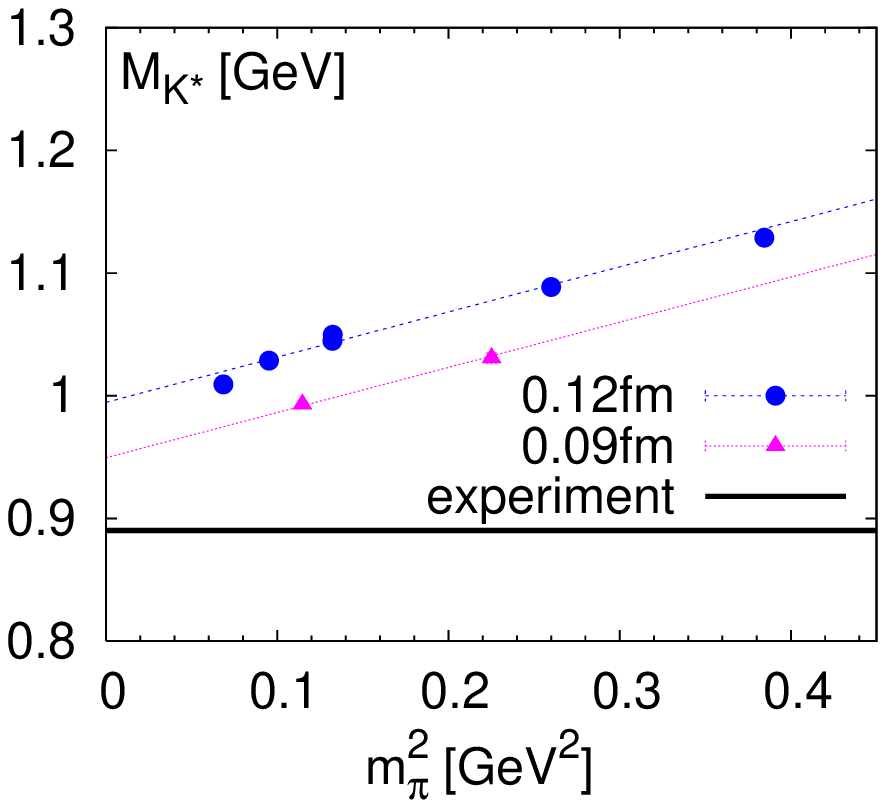}
\includegraphics[width=8cm]{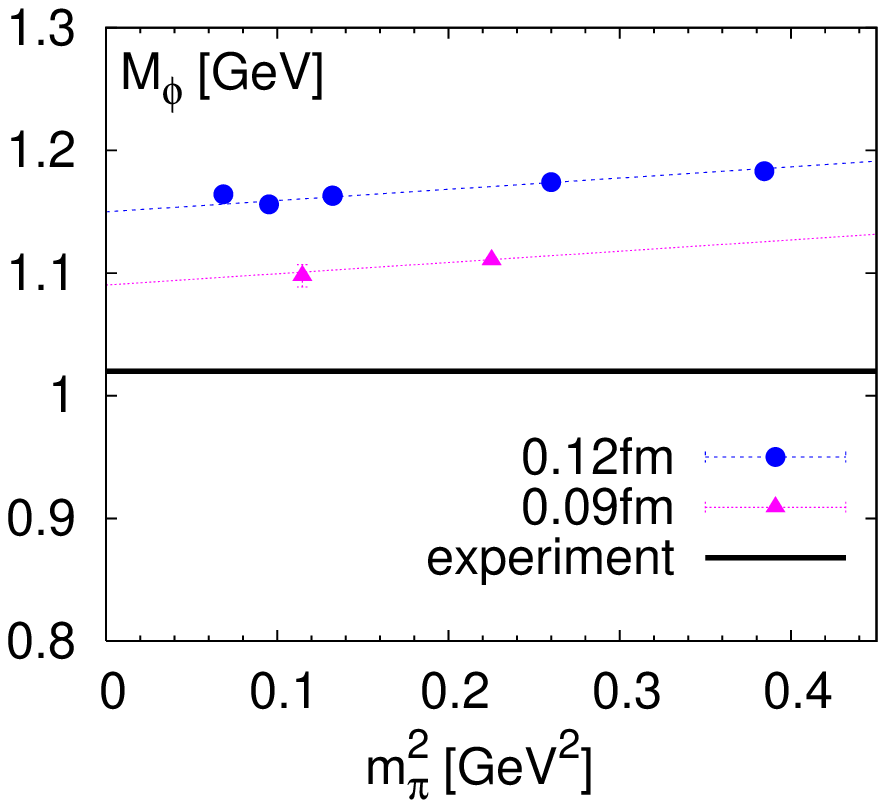}
\caption{The $K^*$ (left) and $\phi$-meson (right) masses calculated on the lattice using
asqtad action \cite{bernard01,bernard04} and compared with Eq. (\ref{fit}) (lines).} 
\label{fig:mVs}
\end{figure}

\begin{figure}
\includegraphics[width=8cm]{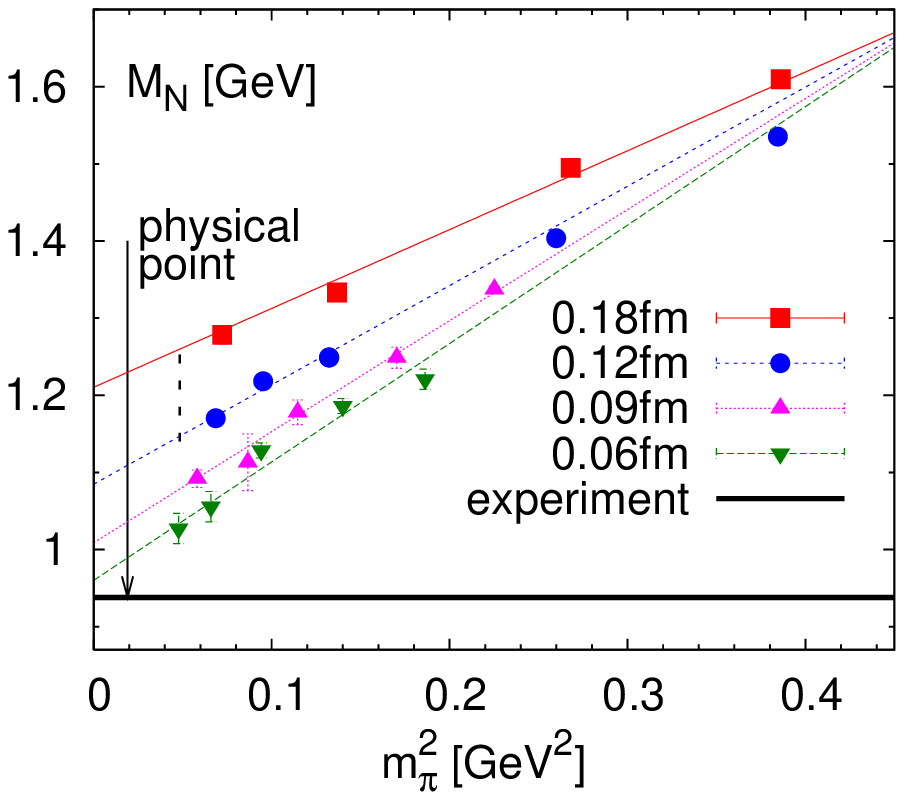}
\includegraphics[width=8cm]{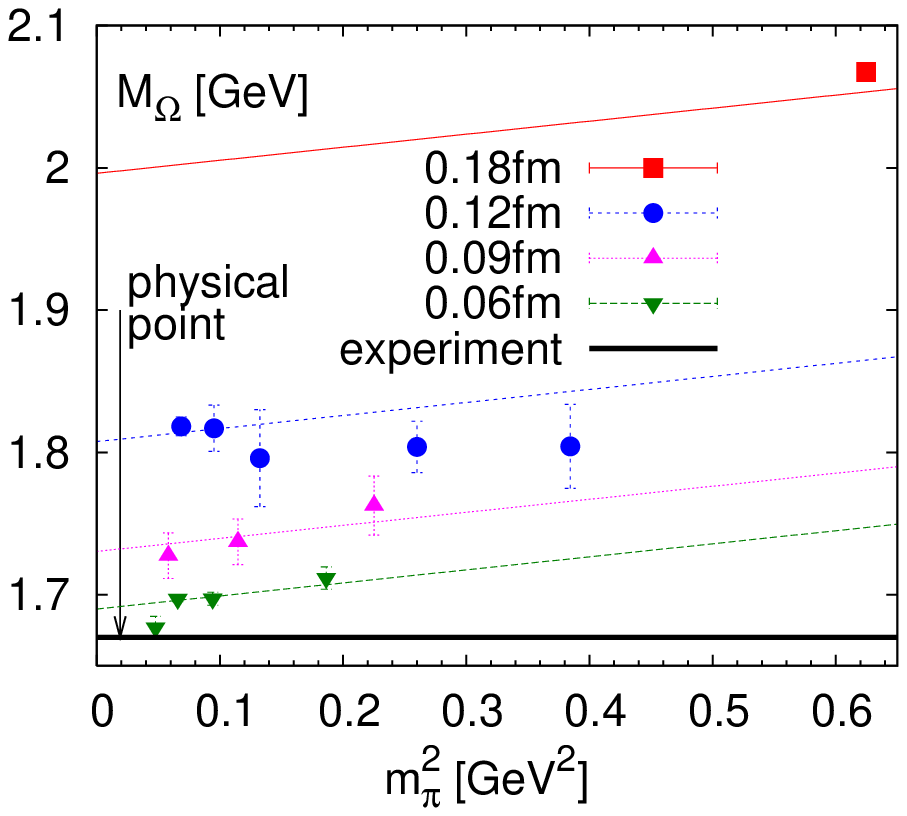}
\caption{The nucleon mass  \cite{bernard01,bernard04,milc_unpub} (left)
and the $\Omega$ baryon mass \cite{bazavov09} (right) calculated with asqtad action and 
compared with our parametrization.}
\label{fig:nucl}
\end{figure}

In this appendix we are going to discuss the cutoff and quark mass dependence of hadron
masses calculated on the lattice and give the parameters entering Eqs. (\ref{mV})-(\ref{mXi}).
We have fitted the quark (pion) mass and lattice spacing dependence of the $\rho,~ K^{*},~\phi,~N$ and $\Omega$ masses
obtained in Refs. \cite{bazavov09,bernard07,bernard04,bernard01} by a simple Ansatz
\begin{equation}
r_1 m= r_1 m_0 + \frac{a_1 (r_1 m_{\pi})^2}{1+a_2 x} + \frac{b_1 x}{1+b_2 x},~x=(a/r_1)^2
\label{fit}
\end{equation}
The values of the fit parameters $m_0$, $a_1,~a_2,~b_1$ and $b_2$ are given in Table \ref{tab:mV}.
In Figs. \ref{fig:mrho}, \ref{fig:mVs} and \ref{fig:nucl} we show the above parametrization 
against the available lattice data. Here we note that the lattice data for the $\Omega$ mass 
have been corrected to take into account that the physical strange quark mass is slightly smaller
that the one used in lattice simulations. 
It turns out that Eq.~(\ref{fit}) reproduces the experimental values of the hadron masses
in continuum limit at the physical point $r_1 m_{\pi}=0.226$. This justifies the use of Eqs. (\ref{mV})-(\ref{mXi})
for the evaluation of the hadron masses in the HRG model. 

\begin{figure}
\hspace*{-0.7cm}
\includegraphics[width=6.3cm]{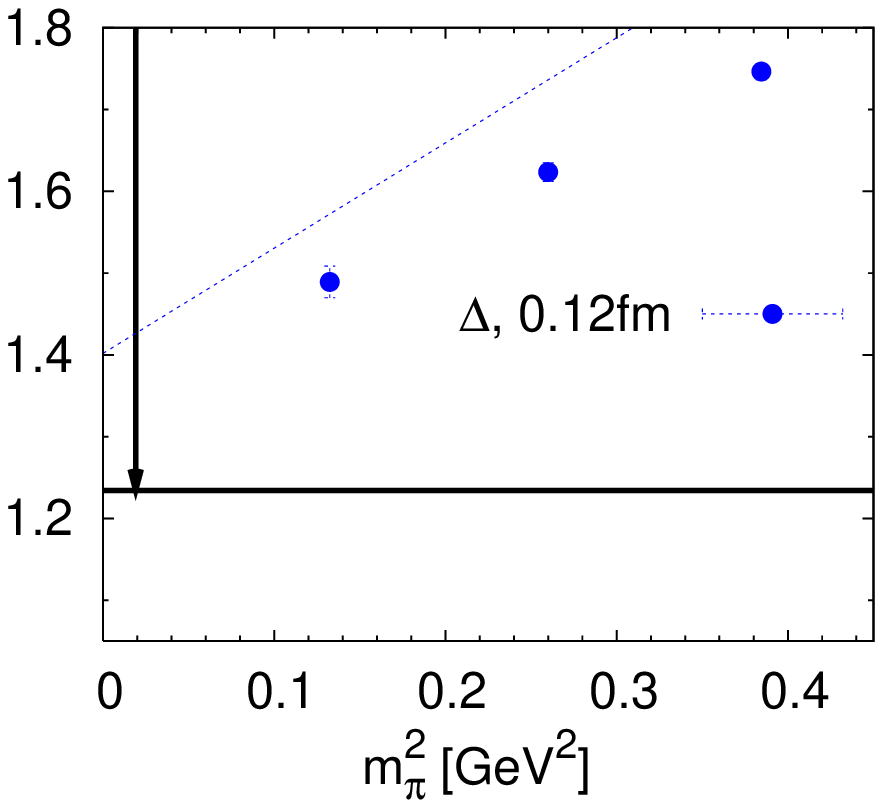}\hspace*{-0.5cm}
\includegraphics[width=6.3cm]{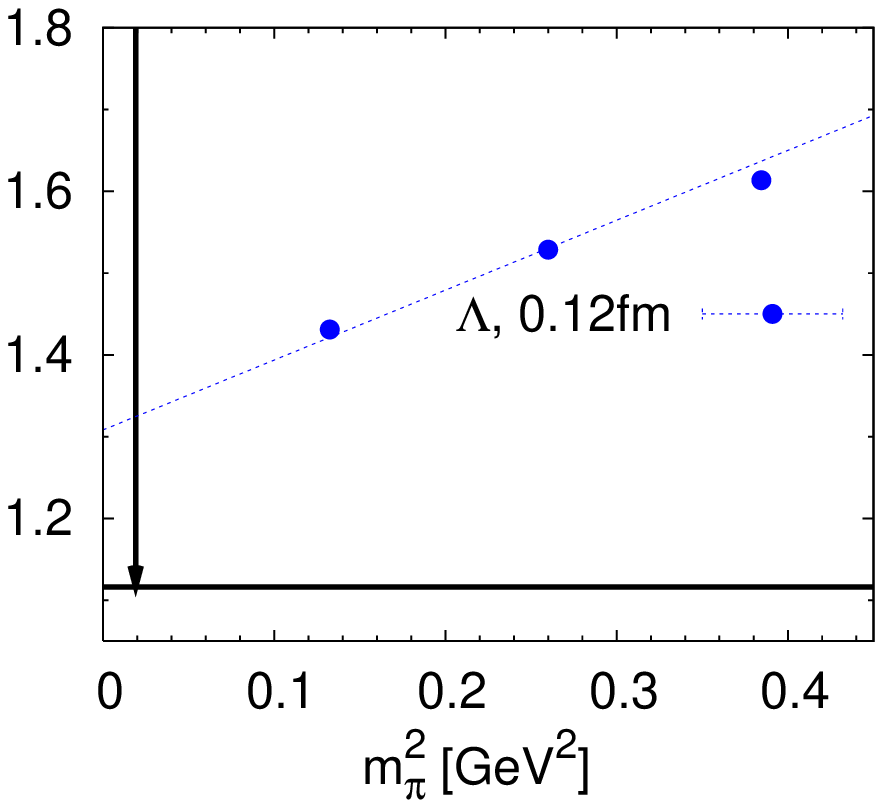}\hspace*{-0.5cm}
\includegraphics[width=6.3cm]{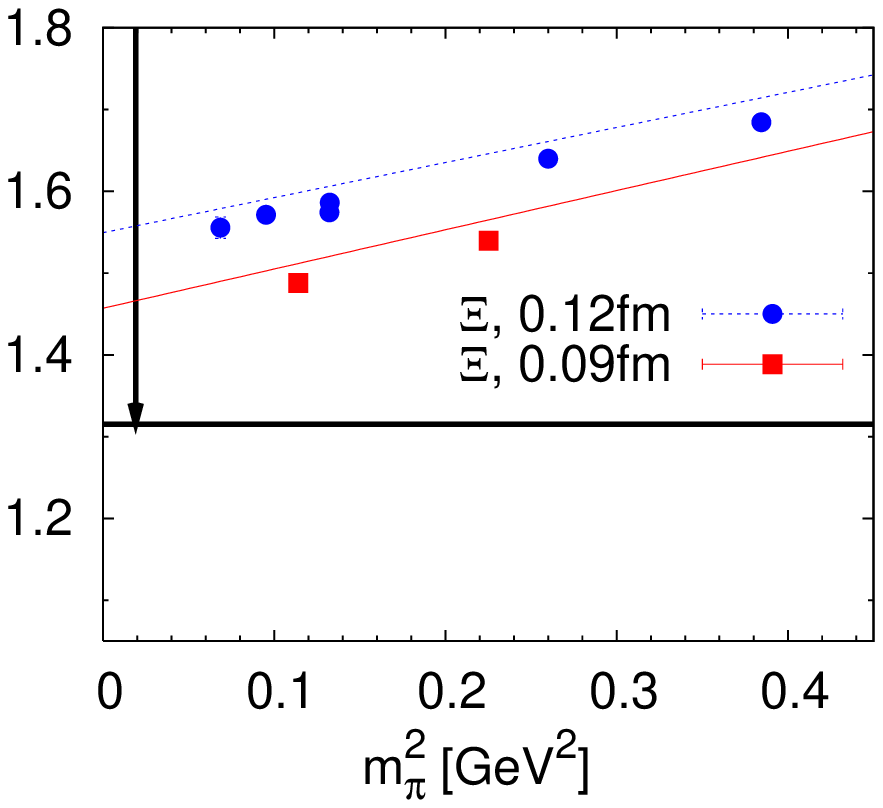}
\caption{The baryon masses calculated for asqtad action \cite{bernard01,bernard04,milc_unpub}
and compared with our parametrization.}
\label{fig:bar}
\end{figure}

As discussed in the main text due to lack of detailed lattice studies we used Eq.~(\ref{mV}) and
Eqs.~(\ref{mLambda})-(\ref{mXi}) to evaluate the mass of the $\Delta$ resonance as well as single
and double strange baryons with values of the parameters in Table \ref{tab:mV} corresponding to 
the nucleon. In Figure \ref{fig:bar} we compare our estimates of the $\Delta$, $\Lambda$ and $\Xi$ baryon
masses shown as lines with available lattice data from the MILC collaboration \cite{bernard04,bernard01}.

\begin{table}
\begin{tabular}{|c|c|c|c|c|c|}
\hline
       &  $r_1 m_0$  & $a_1$    & $b_1$   & $a_2$    & $b_2$   \\
\hline
$\rho$ & 1.17856 & 0.496745 & 2.03538 & 0.958366 & 6.39177 \\
$K^*$  & 1.41505 & 0.228217 & 2.07898 & 0        & 0       \\
$\phi$ & 1.60476 & 0.056901 & 2.74519 & 0        & 0       \\
$N$    & 1.51418 & 0.933405 & 1.65138 & 1.27391 & 1.65138  \\
$\Omega$ & 2.66589 & 0.056779 & 1.74275 & 0       & 0      \\
\hline
\end{tabular}
\caption{The values of the parameters appearing in Eq. (\ref{fit}) for different hadrons.}
\label{tab:mV}
\end{table}


\section{Fitting procedure}
   \label{sec:fit}

Here we describe in detail how we fit the trace anomaly of lattice and
hadron resonance gas. The two first terms of the inverse polynomial
Ansatz (Eq.~\ref{e-3p_high})
\begin{equation}
(e-3P)/T^4 = d_2/T^2 + d_4/T^4 + c_1/T^{n_1} + c_2/T^{n_2}
\end{equation}
appear to provide good fit of the lattice data at
high temperatures, $T>250$ MeV~\cite{hot_eos}. We want to join this
parametrization to the trace anomaly of hadron resonance gas and
require that the trace anomaly and its first and second
derivative with respect to temperature are continuous where
joined. Thus, we need one additional term with negative coefficient
$c_1$ and exponent $n_1>4$ to produce a peak around $T\approx 200$
MeV, and another with positive coefficient $c_2$ and exponent $n_2 >
n_1$ to make the second derivative continuous. We calculate the trace
anomaly of hadron resonance gas using all the resonances up to 2 GeV
mass\footnote{The list of included resonances and their
               properties can be found at~\cite{wiki}.}
in the summary of the 2004 edition of the Review of Particle
Physics~\cite{PDG2004}. We note that including all
the resonances up to $2.5$GeV instead of $2$GeV mass gives
result for the trace anomaly, which is only $2\%$ larger at $T=170$MeV and 
$3\%$ larger at $T=180$MeV. This change is definitely smaller than the expected
discrepancies between HRG and lattice at these temperatures.

In the ansatz we have seven unknown parameters: the coefficients
$d_2$, $d_4$, $c_1$ and $c_2$, exponents $n_1$ and $n_2$, and the
switching temperature $T_0$. We have four constraints, the continuity
of the trace anomaly and its derivatives at $T_0$, and the requirement
$s(T=800\,\mathrm{MeV}) = 0.95\cdot s_{\mathrm{SB}}$ or
$s(T=800\,\mathrm{MeV}) = 0.90\cdot s_{\mathrm{SB}}$.
The requirement
of the continuity of the derivatives gives two equations to fix the
parameters $c_1$ and $c_2$:
\begin{eqnarray*}
c_2 & = & \frac{2(n_1-2)}{n_2(n_2-n_1)}T^{n_2-2}d_2
         +\frac{4(n_1-4)}{n_2(n_2-n_1)}T^{n_2-4}d_4      \\
    & + & \frac{n_1+1}{n_2(n_2-n_1)}T^{n_2+1}G_1(T_0)
         +\frac{1}{n_2(n_2-n_1)}T^{n_2+2}G_2(T_0)         \\
c_1 & = &  \frac{2(n_2-2)}{n_1(n_1-n_2)}T^{n_1-2}d_2
         +\frac{4(n_2-4)}{n_1(n_1-n_2)}T^{n_1-4}d_4       \\
    & + & \frac{n_2+1}{n_1(n_1-n_2)}T^{n_1+1}G_1(T_0)
         +\frac{1}{n_1(n_1-n_2)}T^{n_1+2}G_2(T_0),
\end{eqnarray*}
where
\begin{equation*}
\left.\frac{d}{dT}\frac{\epsilon -3P}{T^4}\right|_{HG} \equiv G_1(T)
\qquad\mathrm{and}\qquad
\left.\frac{d^2}{dT^2}\frac{\epsilon -3P}{T^4}\right|_{HG} \equiv G_2(T).
\end{equation*}

From Eq.(\ref{P-integral}) and $Ts = \epsilon + P$ we obtain 
\begin{equation}
 \frac{s}{T^3} = d_2\left(\frac{2}{T_0^2}-\frac{1}{T^2}\right)
               + \frac{d_4}{T_0^4}
               + \frac{c_1}{n_1}
                 \left(\frac{4}{T_0^{n_1}}+\frac{n_1-4}{T^{n_1}}\right)
               + \frac{c_2}{n_2}
                 \left(\frac{4}{T_0^{n_2}}+\frac{n_2-4}{T^{n_2}}\right)
               +\frac{4P(T_0)}{T_0^4}.
 \label{entropy}
\end{equation}
Since entropy at $T=800$ MeV is fixed, we can use the above equation
to constrain $d_4$,
\begin{equation}
 d_4 = d_4(d_2,n_1,n_2,T_0).
\end{equation}
We can thus express the parameters $c_1$, $c_2$ and $d_4$ in terms of
$d_2$, $n_1$, $n_2$ and $T_0$, and use the continuity of the trace
anomaly to fix $T_0$. We get an equation
\begin{equation}
 \left.\frac{\epsilon -3P}{T^4}\right|_{HG}(T_0)
 = \frac{d_2}{T_0^2} + \frac{d_4(d_2,n_1,n_2,T_0)}{T_0^4}
   +\frac{c_1(d_2,n_1,n_2,T_0)}{T_0^{n_1}}
   +\frac{c_2(d_2,n_1,n_2,T_0)}{T_0^{n_2}},
 \label{continuity}
\end{equation}
which can be evaluated numerically to obtain $T_0$.  This procedure
leaves us with three unknowns $d_2$, $n_1$ and $n_2$, which are chosen
to fit the lattice data. However, such a fitting procedure would be
highly nonlinear. We simplify the problem by requiring that the
exponents are integers, and use brute force: We make a single
parameter ($d_2$) fit with all the integer values $4<n_1<31$, and
$n_1<n_2<43$, and choose the values $n_1$ and $n_2$ which lead to the
smallest $\chi^2$.  Alternatively we can fix $T_0$ to a prescribed
value, use Eq.~(\ref{continuity}) to fix the value of $d_2$, and use
only $n_1$ and $n_2$ to perform the fit.

In our fit we use the lattice data for $T> 250$ MeV obtained with p4 action on $N_{\tau}=8$ lattices
as they extend to sufficiently high temperature \cite{hot_eos}. In addition, we include 
$N_\tau=6$ p4 data for $T>500$MeV \cite{p4_eos}. The fits in general do not reproduce the lattice
data in peak region ($190{\rm MeV} <T< 250 {\rm MeV}$). On the other hand the 
height of the peak in the trace anomaly may be affected by discretization effects. This
can be seen as a difference between the $N_{\tau}=6$ and $N_{\tau}=8$ results.
Assuming that discretization effects in the peak height
go like $1/N_\tau^2$ we can estimate the trace anomaly at $T=206$MeV to
be $5.7 \pm 0.15$. We can use this value as an additional data point in our fits.
We label different parametrization of the trace anomaly 
obtained using different constraints on the entropy density at $T=800$MeV and its height
in the temperature region $190{\rm MeV} <T< 250 {\rm MeV}$
as $s95p\mathrm{-v1}$, $s95n\mathrm{-v1}$, $s90f\mathrm{-v1}$ etc. 
The first three characters stand for the constrain on the entropy density (90\% or 95\% of
the ideal gas value). The fourth character stands for additional constraints on the trace
anomaly in the peak region.
Namely, ``n'' stands for a fit with no constraints: using data for
  $T>250$ MeV, and $T_0$ as a free parameter in the fit. ``p'' means having an additional data 
point of $5.7 \pm 0.15$ at $T=206$ MeV to constrain the peak, and ``f'' stands for a fixed value $T_0=170$ MeV in the fit.
Finally ``v1'' is the version number of the current parametrization.
The value of the parameters $d_2$, $d_4$, $c_1$, $c_2$, $n_1$, $n_2$ and $T_0$
for different fits are given in Table \ref{tab:par}.
This procedure was designed for numerical applications when the trace
anomaly is numerically evaluated using Eq.(\ref{eq:ZHRG}) and laws of
thermodynamics. For practical purposes we also provide a parametrized
version of the trace anomaly of the hadronic part of our EoS. We
choose a polynomial
\begin{equation}
\frac{\epsilon -3P}{T^4} = a_1T^{l_1} +  a_2T^{l_2}
                         + a_3T^{l_3} + a_4T^{l_4}
  \label{HRGpara}
\end{equation}
and fit it to the trace anomaly of the hadron resonance gas evaluated
in the temperature interval $70 < T/\mathrm{MeV} < 190$ with 1 MeV
steps assuming that each point has equal ``error''. The limits have
entirely utilitarian origin: in hydrodynamical applications the system
decouples well above 70 MeV temperature and only a rough approximation of the EoS,
  $P=P(\epsilon)$, is needed at lower temperatures. On the other hand
we expect to switch to the lattice parametrization 
below 190 MeV, and the HRG EoS above that is not needed
either. We fix the exponents in Eq.(\ref{HRGpara}) again using brute
force. We require them to be integers, go through all the combinations
$0\leq l_1<l_2<l_3<l_4\leq 10$, fit the parameters $a_1$, $a_2$,
$a_3$, $a_4$ to the HRG trace anomaly evaluated with 1 MeV intervals,
and choose the values $l_1,~l_2,~l_3$ and $l_4$ which minimize the $\chi^2$. We end
up with $l_1 = 1$, $l_2 = 3$, $l_3 = 4$, $l_4 = 10$, and $a_1 =
4.654$ GeV$^{-1}$, $a_2 = -879$ GeV$^{-3}$, $a_3 = 8081$ GeV$^{-4}$,
$a_4 = -7039000$ GeV$^{-10}$ .
To obtain the EoS, one also needs the pressure at the lower
limit of the integration  (see Eq.(\ref{P-integral}))
$T_\mathrm{low} = 0.07$ GeV: $P(T_\mathrm{low})/T^4_\mathrm{low}=0.1661$.
Our EoSs are also available in a tabulated form at~\cite{wiki}.

\section{Spectra and elliptic flow for partial chemical equilibrium}
\label{sec:PCE}

\begin{figure}
\includegraphics[width=8.7cm]{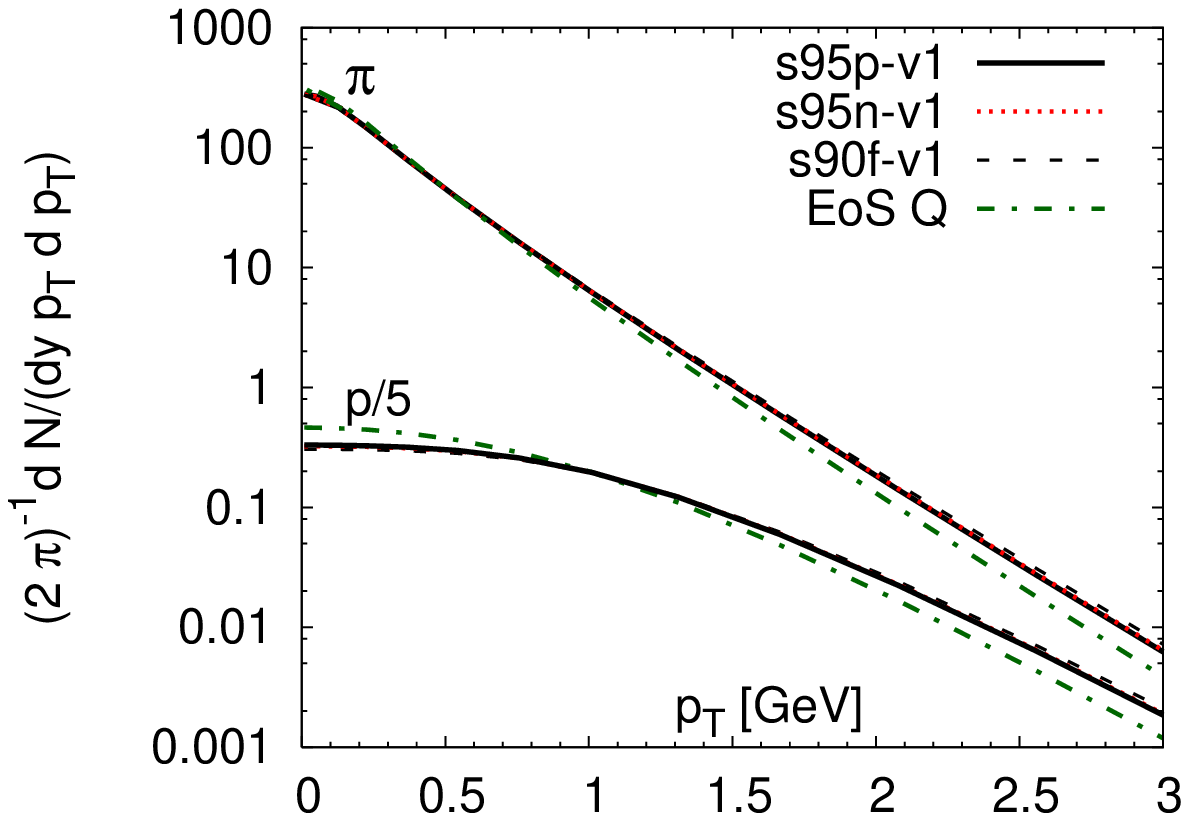}
\includegraphics[width=7.0cm]{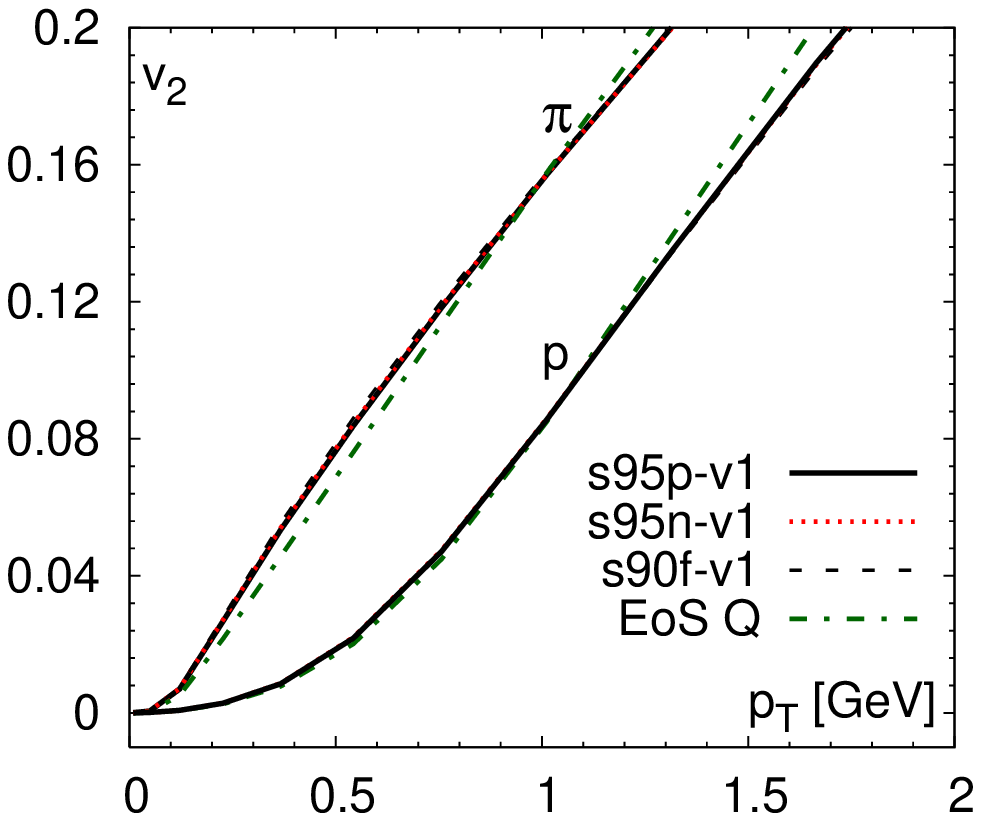}
\caption{The $p_T$-differential elliptic flow $v_2(p_T)$ of protons
  and pions (left) and proton and pion spectra (right) for different
  EoSs in $b=7$ fm Au+Au collision when chemical freeze-out takes
  place at $T_\mathrm{chem}=150$ MeV and kinetic at
  $T_\mathrm{kin}=120$ MeV.}
\label{fig:flow_fixed_ch} 
\end{figure}

In this appendix we discuss the elliptic flow and the spectra of
protons and pions and their sensitivity on EoS when partial chemical
equilibrium~\cite{Bebie:1991,hirano_pce,pasi07} is assumed to
reproduce the observed particle yields. 
The EoS for the system in partial chemical equilibrium is
available in tabulated form [62]. We again calculate the flow in
Au+Au collision at $\sqrt{s_\mathrm{NN}}=200$ GeV with impact
parameter $b=7$ fm.  First we used the same initial condition and the
same freeze-out temperature for all EoSs, namely
$T_\mathrm{chem}=150$MeV for the chemical freeze-out temperature and
$T_\mathrm{kin}=120$MeV for the kinetic freeze-out temperature. The
initial time for the hydrodynamic evolution was chosen to be
$\tau=0.6$fm and the initial entropy density was chosen as for the
case of chemical equilibrium.  The results are shown in Figure
\ref{fig:flow_fixed_ch}. As one can see the elliptic flow is not very
sensitive to the choice of EoS if everything else is kept unchanged,
and all three parametrizations used in the analysis give almost the
same result. The spectra are much more sensitive to the EoS but again
there is no difference for the different lattice parametrizations.

\begin{figure}
\includegraphics[width=8.4cm]{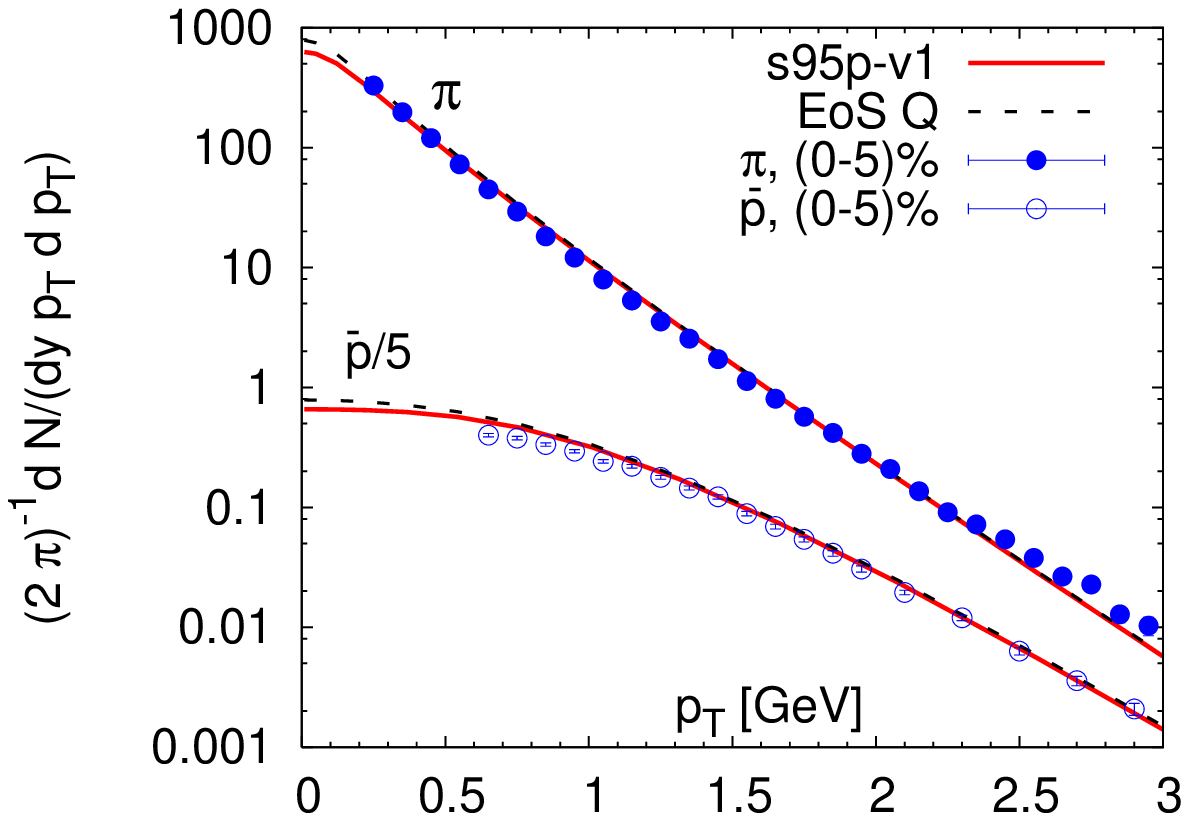}
\includegraphics[width=7.5cm]{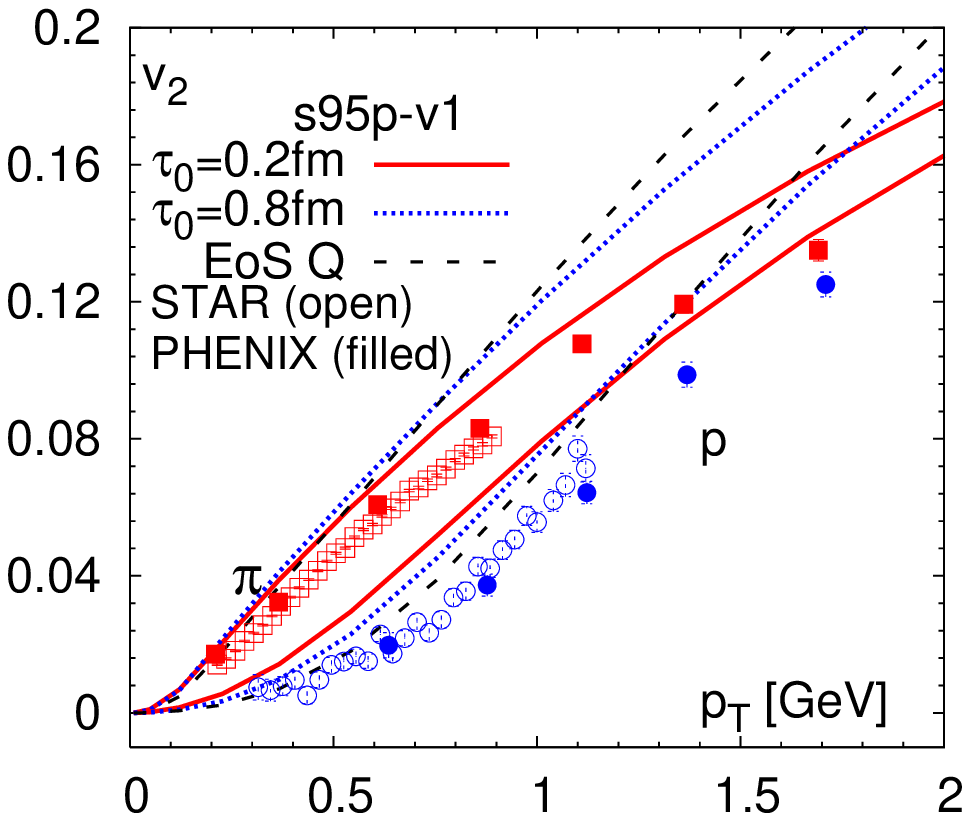}
\caption{Pion ($\pi^+$) and net-proton ($p-\bar{p}$) spectra in 0-5\%
  most central (left), and pion and anti-proton $p_T$-differential
  elliptic flow $v_2(p_T)$ in minimum bias Au+Au collisions at
  $\sqrt{s_\mathrm{NN}}=200$ GeV compared with hydrodynamic
  calculations using two different EoSs and assuming chemical
  freeze-out at $T_\mathrm{chem}=150$MeV. The calculation using EoS
  {\it s95p}-v1 was done using two different initial states, see the text.
  The data was taken by the PHENIX~\cite{Phenix} and the
  STAR~\cite{Star} collaborations. }
\label{fig:flow_fit_ch}
\end{figure}

For a proper discussion on the sensitivity of elliptic flow on the
EoS, one has to again re-tune the calculation to reproduce the
experimental results on the $p_T$-spectra. We obtained the best
results by keeping the chemical and kinetic freeze-out temperatures,
$T_\mathrm{chem}=150$MeV and $T_\mathrm{kin}=120$MeV unchanged, and
tuned the initial conditions.  For EoS\,Q we used $\tau_0=0.2$fm and
initial entropy density which scales with the number of binary
collisions (see Ref.~\cite{pasi07}). For the $s95p$-v1 parametrization
of the EoS we used two different initial conditions. One with
$\tau_0=0.8$fm and initial entropy density proportional to the number
of binary collisions, and a second one with $\tau_0=0.2$fm and initial
entropy density scaling with a combination of binary collisions and
number of participants. The corresponding results are shown in
Figure~\ref{fig:flow_fit_ch}. We see again that the lattice based EoS
give larger $p_T$-differential elliptic flow for the protons than
EoS\,Q for both initial conditions.  In this case, however, the EoS\,Q
does not do a good job of describing the data either, in agreement
with previous findings \cite{hirano_pce,pasi07}. Especially the pion
$p_T$-differential $v_2$ is too large for all EoSs, and there is
clearly room for significant dissipation to reduce the anisotropy. It
is also worth to notice that the uncertainty related to the initial
state is at least as large as the effect of the EoS on the proton
$v_2(p_T)$ and better theoretical constraints to the initial state are
needed.


\begin{thebibliography}{99}
\bibitem{milc_unpub}
Baryon and vector meson masses for $0.15$fm and $0.18$fm were calculated by the
MILC collaboration but not published
\end{thebibliography}
\end{document}